\documentclass[fleqn,usenatbib]{mnras}
\usepackage{lmodern}
\usepackage{newtxtext,newtxmath}

\usepackage[T1]{fontenc}

\DeclareRobustCommand{\VAN}[3]{#2}
\let\VANthebibliography\thebibliography
\def\thebibliography{\DeclareRobustCommand{\VAN}[3]{##3}\VANthebibliography}

\usepackage{graphicx}	
\usepackage{amsmath}	
\usepackage{caption}
\usepackage{subcaption}
\usepackage{comment}
\usepackage{placeins}
\usepackage{makecell}

\newcommand{\galc}[1]{\textcolor{green}{ {GC: #1}}}
\newcommand{\au}{\, {\rm au}}
\newcommand{\me}{\, {\rm M}_{\oplus}}
\newcommand{\Stokes}{{\, {\rm St}}}

\title[External photoevaporation and pebble accretion ]{Formation of multi-planetary systems via pebble accretion in externally photoevaporating discs in stellar clusters}

\author[L. Qiao et al.]{
Lin Qiao,$^{1}$\thanks{E-mail: lin.qiao@qmul.ac.uk}
Gavin A. L. Coleman,$^{1}$
Thomas J. Haworth$^{1}$
\\
$^{1}$Astronomy Unit, School of Physics and Astronomy, Queen Mary University of London, London E1 4NS, UK\\
}

\date{Accepted XXX. Received YYY; in original form ZZZ}

\pubyear{2021}

\begin{document}
\label{firstpage}

\pagerange{\pageref{firstpage}--\pageref{lastpage}}
\maketitle

\begin{abstract}
In this paper, we investigate how external photo-evaporation influences the formation, dynamical evolution and the resultant planetary architecture of multi-planet systems born in stellar clusters. We use a model of N-body simulations of multiple planet formation via pebble accretion coupled with a 1-D viscous disc subject to external photo-evaporation. We found that external photo-evaporation reduces the planet growth by reducing the pebble mass reservoir in discs containing multiple planetary embryos across a wide range of disc masses, and is particularly effective in suppressing planet growth in less initially massive discs (< 0.1\,M$_{\odot}$). However, in more initially massive ($\geq$ 0.1\,M$_{\oplus}$) discs planets lost due to planet-planet interactions dominate the outcome for final resultant total planet mass, masking the effects of external photo-evaporation in curbing the planet mass growth.  In terms of the final resulting planetary architectures, the signature of external photo-evaporation is visible in less massive (< 0.1\,M$_{\odot}$) discs, with fewer numbers and lower masses of planets surviving in discs irradiated with stronger external FUV radiation. External photo-evaporation also leaves a signature for the wide orbit (> 10\,au) terrestrial planets (0.1 - 1\,M$_{\oplus}$), with fewer planets populating this region for stronger FUV field. Finally, the 1st-order resonant pairs fraction decreases with stronger FUV radiation, although the resonant pairs occur rarely regardless of the FUV radiation environment, due to the small number of planets that survive gravitational encounters. 
\end{abstract}




\section{Introduction}
The rapid progress made in today's observational technology has facilitated the discovery of thousands of exoplanets and exoplanet systems. The ever growing population of exoplanets is revealing a wide range of demographics and architectures which are quite different from the Solar System \citep[for detailed
information see reviews by e.g.][]{2021exbi.book....2G, 2015ARA&A..53..409W, 2021ARA&A..59..291Z, 2022cosp...44..571W, 2023ASPC..534..839L}. Thanks to many large surveys with facilities such as Kepler \citep{2010Sci...327..977B}, TESS \citep{2015JATIS...1a4003R} and the HARPS spectrograph \citep{2012SPIE.8446E..1VC, 2013A&A...549A.109B}, the statistics from the exoplanets revolution enables us to search for trends in the parameters of exoplanets and their host stars, which provide valuable constraints on the theories of planet formation and evolution \citep[see reviews by e.g.][]{2022arXiv220309759D, 2022ASSL..466....3R,2023EPJP..138..181E}. Furthermore, the observation of the protoplanetary discs, thanks to the high resolution, high sensitivity observations with facilities such as VLT/SPHERE and ALMA, provides important constraints on the theories of planet formation and disc-planet interaction \citep[for an overview of constraints from disc observations see e.g.][]{2020ARA&A..58..483A, 2023ASPC..534..501M, 2023ASPC..534..539M}

One of the eminent theories of planet formation is core accretion \citep{Pollack,Kokubo96}. In the current theories of core accretion scenario for planet formation, the relatively newly proposed pebble accretion mechanism has been increasingly recognised as an efficient mechanism of forming giant planets by resolving two major obstacles of the traditional core accretion model \citep{2012A&A...544A..32L}. The first issue concerns the planetesimal accretion rates of the planetesimal hypothesis. With planetesimal accretion alone, in order to form the core masses of Jupiter and Saturn, very high planetesimal column densities are needed inside their feeding zones \citep{Pollack,2014MNRAS.445..479C,ColemanNelson16b}. The core masses of Uranus and Neptune are also difficult to form as the planetesimal accretion rates decrease with the orbital distances from the host star \citep{1969Icar...10..109S}. The second issue is when cores grow to about 1 M$_{\oplus}$ in the outer disc region, the assumption that they continue to accrete planetesimals of small velocity dispersion and in isolation from other cores is not valid \citep{2010AJ....139.1297L}. Dynamical simulations have shown that instead of accreting planetesimals, the encounters between cores and planetesimals leads to scattering, clearing the cores' feeding zones and limiting their growth. These issues can be resolved by the pebble accretion model, in which mm-cm sized pebbles are mainly responsible for planetary growth by their accretion onto larger planetesimals and cores \citep{2012A&A...544A..32L,2014A&A...572A.107L}. Pebbles are accreted on to growing planetary embryos at rates that are several orders of magnitude higher than planetesimal accretion due to the greater kinetic energy dissipation by gas drag when pebbles enter the gravitational reach of a core \citep{2010A&A...520A..43O, 2012A&A...544A..32L, 2010MNRAS.404..475J}.

In order to understand the diversity of exoplanet properties and architectures from the theories of planet formation, the effects of the planets' birth environments should not be neglected. Planets form in discs around young (typically $\lesssim
$ 10\,Myr) stars, which means that planets form over similar timescales as the period over which their young host stars occupy the collapsing star forming regions with enhanced local stellar density compared to the galactic field \citep[e.g. ][]{1978PASP...90..506M, 2003ARA&A..41...57L}. Star-disc systems formed in such stellar clusters with high density of gas and neighbouring stars can be affected via gravitational perturbation from stellar encounters \citep[see e.g.][]{2023EPJP..138...11C}, accretion from the surrounding interstellar medium \citep[e.g.][]{2020NatAs...4.1158P}, and material loss from a thermal wind launched from the outer disc by the external UV irradiation from massive OB stars in the star forming region \citep[for a recent review see][]{2022EPJP..137.1132W}. This latter process is called external photo-evaporation, it depletes the disc mass and causes radius truncation if material from the outer disc is removed by the wind faster than its resupply via viscous spreading, then the disc can be truncated \citep{2007MNRAS.376.1350C, 2017MNRAS.468.1631R, 2018ApJ...860...77E,2022MNRAS.514.2315C,Coleman24MHD}. This paper investigates the role of external photo-evaporation on shaping the final planetary system formed via pebble accretion in such stellar clusters. 

There has been sufficient evidence from both observations and theoretical research that external photo-evaporation can significantly affect disc evolution and population in stellar clusters. Externally photo-evaporating discs have been directly observed as proplyds, which are discs enshrouded in a cometary wind with a visible layer of ionised cusp towards the radiation source, and a tail pointing away from the source. So far they are most commonly detected in the Orion Nebula Cluster (ONC) region, which has an age of $\sim 1 - 3$\,Myr, with the main UV source being the O6V star $\theta^1$C \citep[e.g.][]{1994ApJ...436..194O,1999AJ....118.2350H}, also in other star forming regions, e.g. in NGC 1977 near the B1V star as the main radiation source \citep{2016ApJ...826L..15K, 2012ApJ...756..137B, 2025MNRAS.540.1202C}, and in NGC 2024 \citep{2021MNRAS.501.3502H} where both an O8V and a B star act as the radiation source producing proplyds. One sub-region in NGC 2024 where proplyds were observed has a
very young age of $\sim 0.2 - 0.5$\,Myr, implying that external photo-evaporation can compete with even the earliest evidence for planet formation \citep{2018ApJ...857...18S, 2020Natur.586..228S}. So far proplyds have been detected in a wide range of radiation environments, from FUV strength of 100\,$\textrm{G}_0$\footnote{$\textrm{G}_0$ is the \cite{Habing1968} unit of the FUV radiation field, normalised to 1 in the solar neighbourhood.} \citep{2012ApJ...757...78M} to $> 10^5 \textrm{G}_0$, with estimated mass loss rates $> 10^{-7}$ $\textrm{M}_{\odot} \textrm{yr}^{-1}$ and even $> 10^{-6}$\,$\textrm{M}_{\odot} \textrm{yr}^{-1}$ \citep[e.g.][]{1998AJ....116..322H, 1999AJ....118.2350H, 2002ApJ...566..315H, 2021MNRAS.501.3502H}. They are also observed around a wide range of host masses down to even sub-stellar masses (< 15\,$\textrm{M}_{\textrm{Jup}}$) up to solar mass \citep{2010AJ....139..950R, 2016ApJ...826L..15K, 2016ApJ...833L..16F, 2022MNRAS.512.2594H}. Apart from direct observations, disc surveys across many star forming regions also provide evidence of external photo-evaporation shaping disc populations in terms of decreasing inner disc fractions \citep[or disc lifetimes, e.g][]{2016arXiv160501773G, 2012A&A...539A.119F}, decreasing dust mass with higher UV radiation, e.g. in the ONC \citep[][]{2014ApJ...784...82M, 2018ApJ...860...77E, 2019A&A...628A..85V}, NGC 2024 \citep[][]{2020A&A...640A..27V}, and even in intermediate ($10 - 1000$\,$\textrm{G}_0$) FUV radiation fields in the L1641 and L1647 regions of the Orion A cloud from the SODA survey \citep{2023A&A...673L...2V}. Recent work has shown that trends of decreasing dust mass with higher UV radiation varies significantly across stellar mass, the strength of angular momentum transport in the disc, and the age of the clusters themselves (Coleman \& Van Terwisga subm.). 

Apart from observations, recent progress in theoretical modelling have proven the importance of external photo-evaporation in affecting disc evolution processes. In recent years, with the radiation hydrodynamical models of external photo-evaporation by \citet{2018MNRAS.481..452H, 2023MNRAS.526.4315H} providing a public grid (the {\sc fried} grid) of mass loss rates as a function of disc parameters and FUV radiation field strengths, many research have modelled externally photo-evaporating discs with constant UV field \citep[e.g.][]{2018MNRAS.481..452H, 2019MNRAS.485.1489W, 2019MNRAS.490.5478W, 2020MNRAS.497L..40W, 2020MNRAS.491..903W, 2020MNRAS.492.1279S, 2022MNRAS.514.2315C, 2024A&A...681A..84G, Coleman24MHD, Coleman25, 2023A&A...674A.165W}, and with time varying FUV field obtained from cluster simulations \citep[e.g.][]{2019MNRAS.490.5678C, 2021MNRAS.501.1782C, 2023MNRAS.520.6159C, 2021ApJ...913...95P,2021MNRAS.502.2665P, 2022MNRAS.512.3788Q, 2023MNRAS.520.5331W}. 


There has also been recent research coupling models of planet formation \citep[][]{2023MNRAS.522.1939Q, 2024A&A...689A.338H}, migration \citep[][]{2022MNRAS.515.4287W}, dynamical evolution \citep[][]{2022MNRAS.517.2103D} and planet population synthesis \citep[][]{2024A&A...689A.338H, 2023EPJP..138..181E, 2024MNRAS.530..630C, 2024MNRAS.527..414C} with external photo-evaporation. \cite{2022MNRAS.515.4287W} found that the gas accretion and migration of wide-orbit giant planets in a disc can be suppressed by FUV-induced photo-evaporation. \citet{2023MNRAS.522.1939Q} demonstrated that planet formation via pebble accretion is sensitive to external photo-evaporation of the outer disc. Small dust grains are entrained in the wind \citep{2016MNRAS.457.3593F, 2025MNRAS.539.1414P} and so the fast disc truncation by external photo-evaporation can effectively curb planet growth via pebble accretion by limiting the available pebble mass reservoir in the disc. If a disc born in the stellar cluster is initially embedded in a dense star-forming cloud region, hence is shielded from external photo-evaporation for a period before the cloud disperses, then even a short shielding time can be effective in allowing planet growth compared to discs that are exposed to high UV radiation straight from birth \citep{2023MNRAS.522.1939Q}. \citet{2024A&A...689A.338H} also investigated the effects of external photo-evaporation on planet population formed via pebble accretion by coupling a star cluster formation and feedback simulations and a planet population synthesis model, and found that planetary systems born around low mass ($\lesssim 0.2$ M$_\odot$) stars in a clustered environment tend to have fewer cold Jupiters but more cold Neptunes compared to the population born in an isolated environment. 

Most of the research summarised above have found that external photo-evaporation affects the outcomes of planet formation simulations, but so far they have mainly used simplified models consisting of only one planet per system. However, in reality, planetary systems form in protoplanetary discs with multitudes of planetary embryos/cores which dynamically evolve as a result of planet-planet interactions between each other and planet-disc interactions. These effects can not be captured by modelling of an isolated planetary embryo, and can affect pebble accretion rates onto the planetary embryos/cores, complicating the way in which external photo-evaporation affects the planet growth potential. Therefore, it is necessary to model planet formation in discs that contain multiple planetary embryos to understand how external FUV radiation sculpts the planetary architectures and planetary demographics in stellar cluster environments. In this paper, we aim to investigate how the formation of multiple planetary systems via pebble accretion is affected by external photo-evaporation in a stellar cluster environment. We use a model of N-body simulations of multiple planet formation via pebble accretion coupled with a 1-D viscous disc subject to external photo-evaporation. Compared to the single-core per disc model in the previous research, the effects of gravitational N-body interaction between planetary embryos/cores are included, so that the planets are subject to scattering, dynamical ejections and collisional merging. The effects of external photo-evaporation and cloud shielding on the planet growth via pebble accretion and the resulting planetary architecture are explored.

\section{Numerical Method}
We investigate the formation of multi-planetary systems via pebble accretion in externally photoevaporating protoplanetary discs. We use the N-body simulations of of planet formation coupled with a 1-D model of viscously evolving disc that is subject to external photo-evaporation. The N-body simulations were performed using the Mercury-6 sympletic integrator \citep{1999MNRAS.304..793C}, which computes the dynamical evolution of the planetary embryos/planets, accounting for their gravitational interactions between each other and the central star. Three types of consequences can happen to the planetary embryos/planets as a result of the planet-planet interactions: scattering, ejection and collision. Note that collisions between two bodies are assumed to be completely inelastic, resulting in a single merged body containing both of the colliding masses. This collisional merging encourages core mass growth of planets and can further enhance the pebble accretion rate. The simulation also incorporates prescriptions for type I and type II migration and gas accretion onto planetary cores.

%

\begin{table}
 \centering
 \begin{tabular}{|c|c|c|} \hline 
  {Parameter}  &  {Set 1} &  {Set 2} \\ \hline 
\makecell{ $F_{\textrm{FUV, max}}$ \\ (G$_0$) } & \makecell{ $\{ 10^1, 10^2, 10^3,$ \\ $10^{3.5},  10^4, 10^{4.5}, 10^5\}$ } & $\{10^1, 10^3, 10^5 \}$ \\ \hline
 \makecell{$t_{\textrm{sh}}$ \\ (Myr)} & \{0, 0.1, ... , 1.4, 1.5\} & 0  \\ \hline
 \makecell{M$_{\textrm{d, init}}$ \\ (M$_{*}$) } & 0.1   & \{0.01, 0.02, ..., 0.14, 0.15\} \\ \hline 

  \end{tabular}
  \caption{Summary of simulation values used for the maximum FUV field (Upper row), shielding timescale before irradiation by that FUV field (middle row) and the initial disc mass (lower row). }
 \label{table_simparams}
\end{table}

\subsection{Disc Evolution Model}
\label{subsect_discevol}
We use a 1D viscous disc model where the gas surface density is evolved by solving the standard diffusion equation with an additional term to describe mass loss by external photo-evaporation:
\begin{equation}
\dot{\Sigma}(r) = \frac{1}{r}\frac{d}{dr}\left[ 3 r^{1/2}\frac{d}{dr}(\nu \Sigma r^{1/2}) - \frac{2 \Lambda \Sigma r^{3/2}}{G M_{*}} \right] - \dot{\Sigma}_{PE}(r),
\label{eqn:Surface_dens}
\end{equation}
where $\dot{\Sigma}_{PE}(r)$ is the rate of change of surface density, $\Sigma$, at some radial distance $r$ due to external photo-evaporative wind induced by FUV radiation (see section \ref{external_evap}), and $\Lambda$ is the disc planet torque for gap opening in the disc when the planet gets massive enough, defined as

\begin{equation}
\label{disc_planet_torque}
   \Lambda = \textrm{sin}(r - r_p) q^2 \frac{G M_{*}}{2 r} {\left( \frac{r}{ \left| \Delta_{p} \right|} \right)}^{4} ,
\end{equation}
where q is the planet/star mass ratio, $r_p$ is the planet orbital radius. $\left| \Delta_{p} \right|$ = max($H, \left| r - r_p \right|$) with $H$ being the local disc scale height. We use the standard $\alpha$ model for the disc viscosity \citep{1973A&A....24..337S}
\begin{equation}
    \nu = \alpha {c_{s}}^{2} / \Omega, 
\end{equation}
where $c_s$ is the local sound speed, $\Omega$ is the angular velocity and $\alpha$ is the viscosity parameter, which is set at $\alpha = 10^{-3}$ for all simulations. 

For the initial gas surface density, we use the similarity solution of \citet{1974MNRAS.168..603L}

\begin{equation}
\label{sigma_initial}
    \Sigma = \Sigma_{0}\left(\frac{r}{r_c}\right)^{-1}\exp{\left(-\frac{r}{r_C}\right)},
\end{equation}
where $\Sigma_0$ is the normalisation constant set by the total disc mass (for a given $r_c$), and $r_c$ is the scale radius, which sets the initial disc size. We use a host star mass of M$_{*}$ = 1\,M$_\odot$ for all simulations, and for the set of simulations with constant initial disc mass (set 1), it is set as 0.1\,M$_{*}$. For the set of simulations with varying initial disc mass (set 2), it varies between 0.01\,M$_{*}$ - 0.15\,M$_{*}$. (see table \ref{table_simparams} for a summary of parameter values used in our models.) We set the initial disc size via using an initial scale radius value of $r_{C}$ = 50 au. The inner disc edge location $r_{\textrm{in}}$ is set at 0.05\,au. We use an explicit finite difference scheme for solving equation \ref{eqn:Surface_dens} with a non-uniform mesh grid over 2000 grid cells, for which the grid spacing $\Delta r$ scales with radius \citep{2014MNRAS.445..479C}.  {We set a constant solids-to-gas ratio ($Z_0$) of 0.01 throughout the disc, with solids including both dust particles (with dust-to-gas ratio as $Z_{\textrm{dust}}$) that contribute to the disc opacity and pebbles (with pebbles-to-gas ratio as $Z_{\textrm{peb}}$) produced from dust coagulation: $Z_{0} = Z_{\textrm{peb}} + Z_{\textrm{dust}}$. $Z_{0}$ stays constant, but $Z_{\textrm{peb}}$ and 
$Z_{\textrm{dust}}$ evolve as dust gets converted to pebbles in discs (see section \ref{sub_sec_pebacc} for details).}

For the disc initial temperature, we use a radial profile as:
\begin{equation}
    T_\textrm{{initial}} = T_{\textrm{1au}}\left(\frac{r}{{1\rm au}}\right)^{-0.5}, 
\end{equation}
with $T_{\textrm{1au}}$ = $280K$. The disc temperature is updated after each time step via iteratively solving the following thermal equilibrium equation that balances heating from central star irradiation, heating from residual molecular cloud, viscous heating, and blackbody cooling:
\begin{equation}
    Q_\textrm{{irr}} + Q_{\nu} + Q_\textrm{{cloud}} - Q_\textrm{{cool}} = 0,
\end{equation}
where $Q_{\textrm{irr}}$ is the radiative heating rate due to the central star, $Q_{\nu}$ is the viscous heating rate per unit area of the disc, $Q_{\textrm{cloud}}$ is the radiative heating due to the residual molecular cloud (with temperature 10\,K), and $Q_{\textrm{cool}}$ is the blackbody radiative cooling rate \citep[see equations 4-10 in][for further descriptions of the expressions in calculating the temperature]{2021MNRAS.506.3596C}. 

\subsubsection{Mass loss via external photo-evaporation}
\label{external_evap}
\begin{figure}
    \centering	\includegraphics[width=\columnwidth]{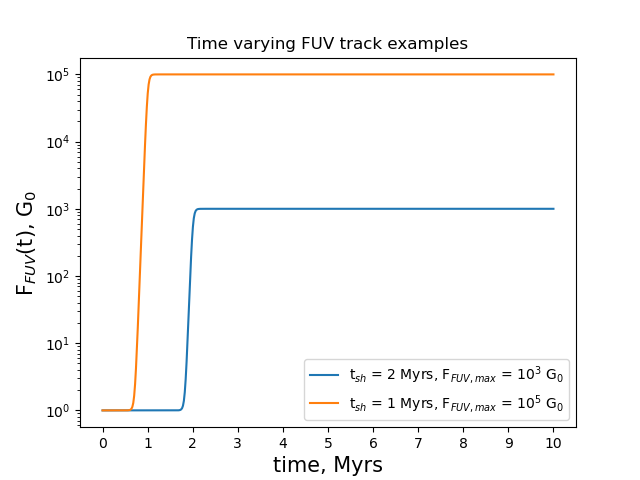}
    \caption{Two examples of some of the time varying FUV tracks implemented the simulation described by equation \ref{eqn:FUV_track}. The blue curve shows an FUV field track with a shielding time $t_{\textrm{sh}} = $ 2 Myrs and a maximum FUV field strength ${F_{\textrm{FUV, max}}} = $ $10^3$ G$_0$. The orange curve shows an FUV field track with a shielding time $t_{\textrm{sh}} = $ 1 Myrs and a maximum FUV field strength ${F_{\textrm{FUV, max}}} = $ $10^5$ G$_0$.} 
    \label{fig:FUV_tracks}
\end{figure}

We implement the external photo-evaporative mass loss from the disc in the same way as \citet{2023MNRAS.522.1939Q}. In order to isolate the impacts of cloud shielding time and FUV field strength on the pebble accretion process and dynamical evolution of the planetary embryos and forming planets in a disc, we use parameterised time varying FUV radiation fields, with shielding time $t_{\textrm{sh}}$ and maximum FUV field strength $F_{\textrm{FUV,max}}$ as free parameters. The FUV field strength starts with a low value of 10 G$_0$ for a period of shielding time $t_{\textrm{sh}}$ to represent the stage when the disc is embedded in the star forming cloud and protected from the strong radiation. After $t_{\textrm{sh}}$ the FUV field strength quickly increases to a constant maximum value $F_{\textrm{FUV,max}}$ representing the stage when the disc is exposed to strong radiation after the cloud dispersal.  {The parameterised FUV radiation as a function of time is mathematically implemented as:
\begin{multline}
\label{eqn:FUV_track}
    F_{\textrm{FUV}}(t) = F_{\textrm{FUV,0}} + \frac{1}{2}(F_{\textrm{FUV,max}}  - F_{\textrm{FUV,0}}) \\ 
    \times \left(\tanh \left(\frac{t - t_\textrm{{shield}}}{t_{\textrm{trans}}}\right) + 1\right),
\end{multline}
where $t_{\textrm{trans}}$ = $5\times10^4$\,years is a parameter that controls the time-scale over which the star/disc moves from a shielded to an unshielded environment. This rapid transition from shielded to irradiated represents the change in irradiation when stars/discs cease being embedded \citep{2022MNRAS.512.3788Q}. Figure \ref{external_evap} shows two examples of the FUV tracks implemented in this paper decribed by equation \ref{eqn:FUV_track}.} Table \ref{table_simparams} summarises the values of $t_{\textrm{sh}}$ and $F_{\textrm{FUV,max}}$ used in the simulations. Note although theoretically both EUV and FUV photons from external massive stars in a cluster can launch photo-evaporative winds from a disc surface, FUV is generally expected to be dominant in setting the mass loss rate \citep{1994ApJ...436..194O, 1998ApJ...499..758J, 2004ApJ...611..360A, 2016MNRAS.457.3593F, 2022EPJP..137.1132W, 2023MNRAS.526.4315H}, therefore we include only FUV radiation to induce external photo-evaporative mass loss in this paper. 

The instantaneous external photo-evaporative mass loss rate at time $t$ from a disc of outer radius $r_d(t)$ and outer surface density $\Sigma_{\textrm{out}}(r_{d}, t)$ induced by the FUV radiation field strength $F_{\textrm{FUV}}(t)$ from the external star at this time is obtained from interpolating {\sc fried} grid v2 \citep{2018MNRAS.481..452H, 2023MNRAS.526.4315H}. The {\sc fried} grid v2 (which is a set of PDR-hydrodynamics simulations of external photo-evaporative mass flows from externally irradiated discs) provides mass loss rates for discs irradiated by FUV radiation as a function of the star/disc/FUV parameters. The {\sc fried} grid contains multiple subgrids that vary the PAH-to-dust ratio ($\rm f_{PAH}$) and specify whether or not grain growth has occurred. The effects of using different combinations of these parameters will be explored in future work, but we do not expect such changes to affect the differences between viscous and MHD wind driven discs. The combination we use here is $\rm f_{PAH}=1$ (an interstellar medium, ISM,-like PAH-to-dust ratio) and assume that grain growth has occurred in the outer disc, depleting it and the wind of small grains which reduces the extinction in the wind and increases the mass loss rate compared to when dust is still ISM-like. This results in a depleted PAH-to-gas ratio of 1/100, which was recommended as fiducial by \citet{2023MNRAS.526.4315H, Coleman25} and is motivated also by \cite{Vicente13} who found signs of PAH depletion in the proplyd HST 10.

At each time step we obtain the mass loss rate via a linear 3D ($r_d(t)$, $\Sigma_{\textrm{out}}(r_{d}, t)$ and $F_{\textrm{FUV}}(t)$) interpolation over the {\sc fried} gird. Special care is taken in defining the outer disc radius location $r_d$, since mass loss rates on the {\sc fried} grid are sensitive to the specific value of disc radius and surface density at this chosen disc edge location. If the disc outer radius is chosen at a location with low surface density which is very optically thin, the mass loss rate interpolated would be unphysical since the FUV radiation would in reality penetrate deeper and drive the wind from the optically thicker region of the disc, resulting in a much higher mass loss rate. This means the disc outer edge location for the purpose of interpolating a physically sensible mass loss rate should be at the optically thick/thin transition, where the interpolated mass loss rate from {\sc fried} is a maximum. We therefore adopt this radius of maximum mass loss $r_{\textrm{max}}$ as the disc outer radius \citep[see ][particularly figure 2 for more information]{2020MNRAS.492.1279S}. At each time step, to find the location of $r_{\textrm{max}}$, we first evaluate the instantaneous mass loss rate at each grid point via interpolating the {\sc fried} grid as if the radius at that grid was the disc outer radius. We then find the location of $r_{\textrm{max}}$ by identifying the grid point location that returns the maximum interpolated rate. With the instantaneous mass loss rate at $r_{\textrm{max}}$ determined, the removal of the external mass loss at each time step is implemented in the same way as detailed in section 2.1.1 of \citet{2023MNRAS.522.1939Q}.

\subsubsection{Magnetopsheric cavity and disc inner edge}
\label{mag_cavity}

In \citet{2023MNRAS.522.1939Q}, as each disc only contained a single planetary embryo, we focused on investigating how the external photo-evaporative mass loss from the disc impacts the planetary growth potential via pebble accretion, without considering the eventual survivability of the planets as the results of migration and planet-planet interactions. In fact, all planets in \citet{2023MNRAS.522.1939Q} simulations that grew to >1 M$_\oplus$ migrated to the inner grid domain boundary (0.1\,au), which was not close enough to the host star to include mechanisms to halt planet migration onto the star. This paper follows previous works \citep{ColemanNelson16,Coleman19} and extends the inner simulation domain to 0.03\,au, so that we can include a magnetospheric disc cavity that extends to the inner edge of the disc at the location of 0.05\,au (this corresponds to an orbital period of $\sim$ 4 days, consistent with the spin periods of many T Tauri stars \citep{2005ApJ...633..967H}. A low density inner magnetospheric cavity will be created if the disc is truncated by a stellar magnetic field, and numerous studies \citep[e.g.][]{1996Natur.380..606L, 2006ApJ...642..478M, 2011A&A...528A...2B, 2011ApJ...741..109T, 2018MNRAS.473.5267M, 2019MNRAS.485.2666R, 2021A&A...648A..69A} have shown that the steep positive gradient at the cavity edge (or the inner edge of the disc) produces a positive one-sided torque which is effective in halting the inward migration of planets. In our simulations,  {in order to simulate the effect of disc inner edge as a planet trap, we stop all type I migration torques on a planet once it reaches the inner edge.} This means a planet that type I migrates inward will stop at the cavity edge, and a planet that grows massive enough to open a gap and type II migrates inward will keep migrating into the cavity until it reaches the 2:1 orbital resonance location with the cavity outer edge (at $\sim$0.0315\,au from the star), at which point disc torques are switched off, halting the migration. It should be noted, that when other subsequent inwardly migrating planets enter into resonance with the halted inner planet, they may be able to push planets closer to star abetting in their destruction, especially
when multiple planets migrate inwards as part of a resonant chain. as was shown in previous work \citep[e.g.][]{ColemanNelson16, Coleman19, 2021A&A...648A..69A}.

\subsection{Pebble Accretion Model}
\label{sub_sec_pebacc}

We take the same approach as \citet{2023MNRAS.522.1939Q} for the model of pebble production and accretion, which is adopted from \citet{2021MNRAS.506.3596C}. This implementation follows the models of \citet{2012A&A...544A..32L} and \citet{2014A&A...572A.107L}, which separate solids into two populations: one of radially stationary small dust grains and one of larger inwardly drifting pebbles. In the model, the small dust grain population grows into larger pebbles via the coagulation of the vertically settling small dust grain population in the disc. The pebble production front $r_g$, which describes the location in the disc at a certain time where the small dust grains have just grown to the pebble size that can drift inwards ($\approx$ 1 - 10mm \citep{2014A&A...572A.107L}, is
\begin{equation}
\label{eqn:pebble_front}
    r_{g}(t) = {\left(\frac{3}{16}\right)}^{1/3}{(G M_{*})}^{1/3}{(\epsilon_{d} Z_{0})}^{2/3}t^{2/3},
\end{equation}
where $\epsilon_{d} = 0.05$ defines the size-dependent dust growth efficiency, and $Z_0 = 0.01$ is the solids-to-gas ratio, contributed by both pebbles ($Z_{\textrm{peb}}$) and dust ($Z_{\textrm{dust}}$):
\begin{equation}
    \label{eqn:Z}
    Z_{0} = Z_{\textrm{peb}} + Z_{\textrm{dust}}.
\end{equation}
In this work we assume that 90\% of the total solids is converted into pebbles, and this conversion rate remains constant throughout the simulation, that is consistent with that found by planet formation models required to form planets through pebble accretion similar to those observed \citep{Brugger20}. As the dust growth time-scale is shorter at smaller disc radii (i.e. dust at smaller radii grows faster), the pebble production front moves outwards from the disc inner edge over time. This provides a pebble mass flux (flux of pebbles drifting inwards from $r_{g}$) defined as
\begin{equation}
\label{eqn:pebble_flux}
    \dot{M}_{\textrm{flux}} = 2 \pi r_g \frac{dr_{g}}{dt} Z_{\textrm{peb}}(r_g) \Sigma_{\textrm{gas}}(r_g),
\end{equation}
where $\Sigma_{\textrm{gas}}(r_g)$ is the gas surface density at the pebble production front. From the mass flux we define the pebble surface density profile $\Sigma_{\textrm{peb}}$ the same way as \citet{2014A&A...572A.107L}
\begin{equation}
    \Sigma_{\textrm{peb}} = \frac{\dot{M}_{\textrm{flux}}}{2 \pi r v_{r}},
\end{equation}
where $r$ is the disc radius and $v_r$ is the radial velocity of pebbles at $r$, defined as
\begin{equation}
    v_r = 2 \frac{\Stokes}{\Stokes^{2} + 1} \eta v_{k} - \frac{v_{\textrm{r,gas}}}{1+ \Stokes^{2}}
\end{equation}
\citep{1977MNRAS.180...57W,1986Icar...67..375N} ,where $\rm St$ is the Stokes number of the pebbles, $v_k$ is the local keplerian velocity, $v_{\textrm{r,gas}}$ is the local gas radial velocity, and $\eta$ is the dimensionless measure of gas pressure support
\begin{equation}
\label{eta}
\eta = -\frac{1}{2}{\left(\frac{H}{r}\right)}^{2}\frac{\partial \textrm{ln}P}{\partial \textrm{ln}r},
\end{equation}
where $H$ is the disc scale height \citep{1986Icar...67..375N}.
For the Stokes number, we assume it is equal to:
\begin{equation}
    \Stokes = \min(\Stokes_{\rm drift}, \Stokes_{\rm frag})
\end{equation}
where $\Stokes_{\rm drift}$ is the drift-limited Stokes number that is obtained through an equilibrium between the drift and growth of pebbles to fit constraints of observations of pebbles in protoplanetary discs and from advanced coagulation models \citep{2012A&A...539A.148B}
\begin{equation}
    \Stokes_{\rm drift} = \dfrac{\sqrt{3}}{8}\dfrac{\epsilon_{\rm p}}{\eta}\dfrac{\Sigma_{\rm peb}}{\Sigma_{\rm gas}},
\end{equation}
where $\epsilon_{\rm p}$ is the coagulation efficiency between pebbles which we assume is similar to the dust growth efficiency and is equal to 0.5.
As well as the drift-limited Stokes number, we also include the fragmentation-limited Stokes number, ($\Stokes_{\rm frag}$) which we follow \citet{Ormel07} and is equal to
\begin{equation}
    \Stokes_{\rm frag} = \dfrac{v_{\rm frag}^2}{3\alpha c_{\rm s}^2}
\end{equation}
where $v_{\rm frag}$ is the impact velocity required for fragmentation, which we model as the smoothed function
\begin{equation}
    \dfrac{v_{\rm frag}}{1 \rm ms^{-1}} = 10^{0.5+0.5\tanh{((r-r_{\rm snow})/5H)}},
\end{equation}
where $r_{\rm snow}$ is the water snowline, which we assume to be where the disc midplane temperature is equal to 170K. The fragmentation velocity therefore varies between $1$\,ms$^{-1}$ for rocky pebbles \citep{Guttler10}, to $10$\,ms$^{-1}$ for icy pebbles, consistent with some results in the literature \citep[though this is still an area of open research][]{Gundlach15,Musiolik19}.  {Note that the stokes number is drift limited in the outer disc, and only fragmentation limited in the inner au or so, and our planet growth via pebble accretion happens in outer disc regions. Therefore the final resultant planet formed is not sensitive to the assumption of the fragmentation velocities for icy and rocky pebbles.}

The pebble accretion rates for a planetary embryo starts with a 3D mode when the embryo's Hill radius is smaller than the pebble scale height:
\begin{equation}
    \dot{M}_{\textrm{3D}} = \pi {R_{\textrm{acc}}}^{2} \rho_{\textrm{peb}} \delta v ,
    \label{mdot_3D}
\end{equation}
where $\rho_{\textrm{peb}}$ is the midplane pebble density, $\delta v=\Delta v + \Omega R_{\textrm{acc}}$ is the approach speed, with $\Delta v$ as the relative sub-Keplerian velocity at which particles approach the planetary embryo/core. When the embryo/planetary core becomes large enough such that its Hill radius is larger than the pebble scale height, the accretion rate switch a 2D mode: 
\begin{equation}
    \dot{M}_{\textrm{2D}} = 2 R_{\textrm{acc}} \Sigma_{\textrm{peb}} \delta v. 
    \label{mdot_2D}
\end{equation}

For the approach speed term $\delta v$ in equation \ref{mdot_3D} and \ref{mdot_2D} is the relative velocity between this embryo and pebbles moving with the sub-Keplerian speed $\Delta v$. When the planetary embryo's orbit has non zero eccentricity and inlination, the relative velocity becomes a function of time defined as:  
\label{v_rel}
\begin{equation}
    v_{\textrm{rel}} = \sqrt{[v_\textrm{e} \cos(\Omega t)]^2 + [-(1/2)v_\textrm{e} \sin(\Omega t) + \Delta v]^2 + [v_\textrm{i} \cos(\Omega t)]^2},
    \label{v_rel2}
\end{equation}
where $v_\textrm{e} = e v_\textrm{K}$ and $v_\textrm{i} = i v_\textrm{K}$ are the eccentricity and inclination speeds of the orbit. The accretion radius, R$_{\textrm{acc}}$, in equation \ref{mdot_3D} and \ref{mdot_2D} also have two regimes: the Bondi regime and the Hill regime. Initially the low mass embryo accretes in the Bondi accretion regime through its Bondi radius $R_B$, which is smaller than its Hill radius $R_H$, but as the embryo (or at this stage the proto-planetary core) becomes more massive such that its Bondi radius becomes similar to its Hill radius, the accretion regimes switches to a Hill regime where the accretion rate is limited by its Hill sphere. $R_{\textrm{acc}}$ depends on both the accretion regime and the pebble friction time-scale. The dependency on friction time arises due to pebbles having to change directions significantly on time-scales shorter than the friction time-scale, which brings a criterion accretion radius $\hat{R}_{\textrm{acc}}$ defined as:
\begin{equation}
    \hat{R}_{\textrm{acc}} = {\left( \frac{4t_f}{t_B} \right)}^{1/2} R_B
\end{equation}
for the Bondi regime, where $t_f = \Stokes/\Omega $ is the pebble friction time-scale, $t_B$ is the Bondi sphere crossing time, and
\begin{equation}
    \hat{R}_{\textrm{acc}} = {\left( \frac{\Omega_{k} t_f}{0.1} \right)}^{1/2} R_H
\end{equation}
for the Hill regime. The accretion radius is the defined as:
\begin{equation}
    R_{\textrm{acc}} = \hat{R}_{\rm acc} \exp{ \left [ - \chi {(t_f/t_p)}^{\gamma} \right ] },
\end{equation}
where $t_p = GM / {(\Delta v + \Omega R_H)}^3$ is the characteristic passing time scale with $\chi = 0.4$ and $\gamma = 0.65$ \citep{2010A&A...520A..43O}. The core mass threshold above which the transition from Bondi accretion to Hill accretion happens is called the transition mass, and is calculated as: 
\begin{equation}
M_{\textrm{trans}} = \eta^{3} M_{*}
\end{equation}
with $\eta$ described by equation \ref{eta}.  

The planetary embryos grow by accreting pebbles until they reach the so-called pebble isolation mass, that is the mass required to perturb the gas pressure gradient in the disc: i.e. the gas velocity becomes super-Keplerian in a narrow ring outside of a planet's orbit reversing the action of the gas drag. The pebbles are therefore pushed outwards rather than inwards and accumulate at the outer edge of this ring stopping the embryos from accreting solids \citep{PaardekooperMellema06,Rice06}. Initial work found that the pebble isolation mass was proportional to the cube of the local gas aspect ratio \citep{2014A&A...572A.107L}. More recent work however has examined what effects disc viscosity and the Stokes number of the pebbles have on the pebble isolation mass, finding that small pebbles that are well coupled to the gas are able to drift past the pressure bump exterior to the planet's orbit \citep{Ataiee18,Bitsch18}. To account for the pebble isolation mass whilst including the effects of turbulence and stokes number, we follow \citet{Bitsch18}, and define a pebble isolation mass-to-star ratio,
\begin{equation}
q_{\rm iso} = \left(q_{\rm iso}^{\dagger} + \frac{\Pi_{\rm crit}}{\lambda} \right) \frac{M_{\oplus}}{M_*}
\end{equation}
where $q_{\rm iso}^\dagger  = 25 f_{\rm fit}$, $\lambda = 0.00476/f_{\rm fit}$, $\Pi_{\rm crit} = \frac{\alpha}{2St}$ and
\begin{equation}
\label{eq:ffit}
 f_{\rm fit} = \left[\frac{H/r}{0.05}\right]^3 \left[ 0.34 \left(\frac{\log(\alpha_3)}{\log(\alpha)}\right)^4 + 0.66 \right] \left[1-\frac{\frac{\partial\ln P}{\partial\ln r } +2.5}{6} \right] \ ,
\end{equation}
with $\alpha_3 = 0.001$.
Once planetary embryos reach the pebble isolation mass, they no longer accrete pebbles from the discs in our simulations.

\subsection{Gas envelope accretion}
\label{sub_sect_gasacc}
Once a proto-planetary core reaches sufficient mass through pebble accretion, it is able to accrete a gaseous envelope. As it is computationally expensive to incorporate 1D envelope structure models \citep[e.g.][]{CPN17} into our simulations, we used gas accretion rate fits provided by \citet{Poon21} based on 1D envelope structure calculations from \citet{CPN17} instead. To calculate these fits, \citet{Poon21} performed numerous simulations, embedding planets with initial core masses between 2--15\,$\me$ at orbital radii spanning 0.2--50\,$\au$, within gas discs of different masses. This allowed for the effects of varying local disc properties to be taken into account when calculating gas accretion fits, a significant improvement on fits from other works \citep[e.g.][]{Hellary,ColemanNelson16}.
With 1D calculations from \citet{CPN17}, the embedded planets were then able to accrete gas from the surrounding gas disc until either the protoplanetary disc dispersed, or the planets reached a critical state where they would then undergo runaway gas accretion.
With the results of these growing planets, \citet{Poon21} calculated fits to the gas accretion rates taking into account both the planet and the local disc properties.

The gas accretion rate from the fits found in \citet{Poon21} that we use is equal to: 
\begin{align}\label{eq:gasenvelope-gc}
\left (\dfrac{d M_{\mathrm{ge}}}{d t} \right )_{\mathrm{local}}=& 10^{-10.199} \left( \dfrac{\mathrm{M_{\oplus}}}{\mathrm{yr}}\right) f_{\mathrm{opa}}^{-0.963} \left ( \dfrac{T_{\mathrm{local}}}{\mathrm{1\,K}}\right )^{-0.7049}\nonumber \\
&\times \left (\dfrac{M_{\mathrm{core}}}{\mathrm{M}_{\oplus}} \right )^{5.6549}  \left (\dfrac{M_{\mathrm{ge}}}{\mathrm{M}_{\oplus}}  \right )^{-1.159}\nonumber \\
&\times\left [ \exp{\left ( \dfrac{M_{\mathrm{ge}}}{M_{\mathrm{core}}} \right )} \right ]^{3.6334}.
\end{align}
where $T_{\rm local}$ is the local disc temperature, $f_{\rm opa}$ is an opacity reduction factor (which reduces the grain opacity contribution, kept constant at $f_{\rm opa} = 0.01$ in all simulations) and $M_{\rm core}$ and $M_{\rm ge}$ are the planet's core and envelope masses respectively.
When comparing the masses of gas accreting planets calculated through eq. \ref{eq:gasenvelope-gc} to the actual masses obtained using the 1D envelope structure model of \citet{CPN17}, \citet{Poon21} found excellent agreement, a considerable improvement on previous fits \citep[e.g.][]{ColemanNelson16}.
In each of our simulations, we allow the planetary embryo to accrete a gaseous envelope once their mass exceeds an Earth mass, and all mass accreted by planets is removed from the disc to conserve mass.

\subsection{Type I migration and Type II migration}
The planetary embryos which grow massive enough (significantly exceeding a Lunar mass) are subject to type I migration, which is implemented in the same way as \citet{2014MNRAS.445..479C}, based on the torque formulae presented by \citet{2010MNRAS.401.1950P,2011MNRAS.410..293P}. For detailed implementation see equation 15 to 24 in section 2.5 in \citet{2014MNRAS.445..479C}. 

Once a type I migrating planet in a disc grows massive enough to reach the gap opening mass, it switches to type II migration. We implement the gap opening criterion (which specifies the minimal mass of a planet to cause a depletion of 90 \% of the gas in the gap of the disc) derived in \citet{2006Icar..181..587C} that balances the gravity of the planet to the viscous and pressure torques:

\begin{equation}
\label{gap_open_criterion}
    \frac{3}{4} \frac{H}{R_H} + \frac{50}{q Re}  \le 1,
\end{equation}

where $q = M_p/M_{*}$, $R_{\textrm{H}} = {(q/3)}^{(1/3)}$ is the planet Hill radius, $Re = {r_p}^2 \Omega_{p} / \nu$ is the Reynolds number of the disc at the location of the planet ($r_{p}$). Once the planet has opened a gap, the type II migration torque per unit mass on the planet is implemented as
\begin{equation}
    \Gamma_{\textrm{II}} = - \frac{2 \pi}{m_p} \int_{r_{\textrm{in}}}^{r_{\textrm{out}}} r \Lambda \Sigma_{g}\,dr ,
\end{equation}
where $\Lambda$ is the disc planet torque per unit mass in equation \ref{disc_planet_torque}.

\subsection{Initial conditions}
In each simulation, the disc contains 64 planetary embryos initially injected between 1 and 40\,au, separated by 10 mutual Hill radii. The initial mass of each planetary embryo core $M_{\textrm{core, init}}$ takes the value of 10\% of the transition mass M$_{\textrm{trans}}$. This is motivated by the study of \citet{2021MNRAS.506.3596C} who investigated the sizes and distributions of planetary embryos and planetesimals formed in discs via pebble trapping in short-lived pressure bumps, and found that the masses of the planetary embryos formed are at least one order of magnitude lower than the transition mass. The initial eccentricities and inclinations for planetary embryos were randomised between 0-0.02 and 0-$0.36^{\circ}$. Planet-planet interactions are chaotic, so for each combination of initial physical parameters (summarised in table \ref{table_simparams}), 5 realisations were run with different random number seeds that set the randomised initial 3D planet positions and velocities. For the purpose of isolating the impact of external photo-evaporation, internal photo-evaporation is not considered. Each simulation is run for a total of 10\,Myr. This accounts for the entirety of the disc lifetimes, as well as a period of N-body evolution in undamped environments.


The effect of external photoevaporation on disc evolution and planet formation is sensitive to the time that the disc is shielded prior to external irradiation \citep{2022MNRAS.512.3788Q}. Therefore, similar to the approach in \citet{2023MNRAS.522.1939Q}, the discs are subject to the external photo-evaporation induced by a parameterised time-varying FUV radiation field \citep[see figure 1 of][]{2023MNRAS.522.1939Q}. Two parameters for the external FUV radiation field are varied: the initial shielding time $t_{\textrm{sh}}$ (0 - 1.5\,Myr, the upper limit of $t_{\textrm{sh}}$ of 1.5\,Myr is motivated by the result in \citet{2023MNRAS.522.1939Q}, which found shielding beyond 1.5\,Myr completely nullifies the impact of FUV radiation for embryos injected at all initial semi-major axes); and the maximum FUV radiation field strength $F_{\textrm{FUV, max}}$ ($10^1 - 10^5$\,G$_0$). This is described further in section \ref{external_evap}.

Two sets of simulations are performed: Set 1 varies $F_{\textrm{FUV, max}}$ and $t_{\textrm{sh}}$ while fixing the initial disc at the fiducial value of $M_{\textrm{d, init}}$ = 0.1 M$_{*}$, thus focusing on investigating the impact of varying FUV radiation strengths ($10^1 - 10^5$ G$_0$) and shielding time (0 - 1.5 Myr) on planet formation and evolution. Set 2 studies how different levels of external radiation impact planet formation as a function of disc mass, varying the initial disc mass $M_{\textrm{d, init}}$ between 0.01 to 0.15 M$_{*}$ and $F_{\textrm{FUV, max}}$. Compared to a finer parameter grid of values of $F_{\textrm{FUV, max}}$ in the range of $10^1 - 10^5 G_0$ in set 1, set 2 uses only three values of $F_{\textrm{FUV, max}}$ $10^1, 10^3, 10^5$, while prioritising a finer parameter grid of values of $M_{\textrm{d, init}}$. Table \ref{table_simparams} summarises the parameter values used in the two simulation sets in this paper.

\section{Influential role of external photoevaporation}

\FloatBarrier
\subsection{Impacts on pebble accretion rates}
\label{planet_forming_potential}

In this section we explore the effects of external photo-evaporative mass loss on affecting the growth potential of planetary embryos via pebble accretion in a multi-planet system. \citet{2023MNRAS.522.1939Q} demonstrated in the case of a single embryo per disc, that rapid radius truncation of the gas disc by external photo-evaporation limits the amount of small dust available to grow to pebble sizes that can drift inward and be accreted onto the embryo. Therefore, just by depleting the gas in the outer disc, the overall pebble mass reservoir in the disc available for the planet core growth via pebble accretion is reduced. The single embryo/planet per disc models in \citet{2023MNRAS.522.1939Q} showed that a sufficiently strong FUV radiation can reduce the final mass of a planet by at least one order of magnitude. Conversely, an initial period of cloud shielding can protect the disc from rapid radius truncation and effectively preserve the pebble reservoir in the outer disc region.

\begin{figure}
    \centering	\includegraphics[width=\columnwidth]{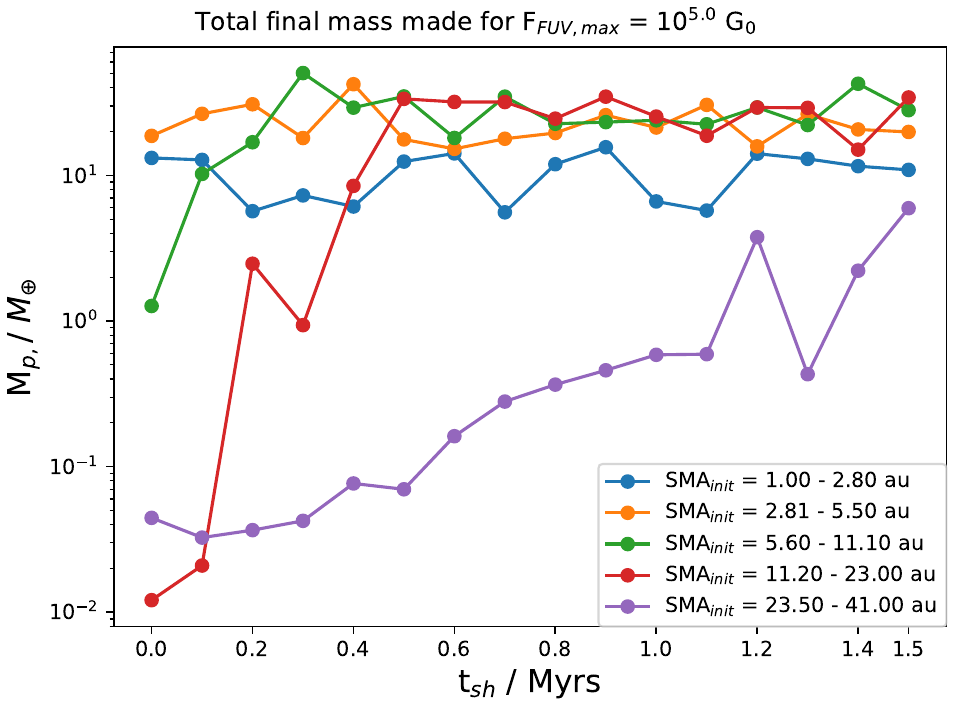}
    \caption{The total final planet mass made per disc (averaged over 5 realisations of the same parameters, counting all planets regardless of whether they are lost due to gravitational interactions) as a function of shielding timescale $t_{\textrm{sh}}$ for planets within various ranges of initial semi-major axis (different colors) in discs irradiated by F$_{\textrm{FUV, max}}$ = $10^5$ G$_0$).} 
    \label{fig:total_m_final_made_selected_ainit_10^5G0}
\end{figure}

The same effect of the overall pebble mass reservoir in a disc being reduced by external photo-evaporation still applies when the disc contains multiple embryos. We therefore still expect the overall growth of the planetary cores to be suppressed by external photo-evaporation and expect cloud shielding to help facilitate planet growth under strong FUV radiation. This is indeed reflected in the final planet masses that embryos in a disc achieved as a function of FUV field strengths F$_{\textrm{FUV,max}}$ and shielding time $t_{\textrm{sh}}$. Figure \ref{fig:total_m_final_made_selected_ainit_10^5G0} shows the sum of the final planet masses of all embryos per disc (averaged over 5 realisations of the same parameters) that originated in a certain range of initial semi-major axis locations (SMA$_{\textrm{init}}$) as a function of shielding time $t_{\textrm{sh}}$. The discs in this plot have the fiducial initial disc mass of 0.1\,M$_{*}$, and are subject to the strongest radiation field of F$_{\textrm{FUV,max}}$ of $10^5$ G$_0$.  For a fair comparison, all of the SMA$_{\textrm{init}}$ ranges (indicated by different colours) contain the same number of planetary embryos. \citet{2023MNRAS.522.1939Q} found that for the single embryo/planet per disc case, there is a non-linear correlation of a higher final mass achieved by the planet with a longer shielding time (up to a certain value), for strongly irradiated discs. From Fig. \ref{fig:total_m_final_made_selected_ainit_10^5G0}, the non-linear relationship between the total M$_{\textrm{final,p}}$ formed in a disc and $t_{\textrm{sh}}$ can still be seen, especially for embryos initially located between 5.6 and 41 au. For embryos in this range, a longer initial cloud shielding period for the disc preserves the pebble mass reservoir and results in more massive planets forming in a disc, up to a certain threshold value of $t_{\textrm{sh}}$. Beyond this $t_{\textrm{sh}}$ value (e.g. $\sim 0.3$\,Myr for embryos initially between 2.8 and 5.5\,au and $\sim 0.5$\,Myr for embryos initially between 5.6 and 11.1\,au), the shielding time is long enough for the embryos to have enough accretion time to reach the pebble isolation mass before exposing the disc to strong external photo-evaporation, therefore further increasing the shielding time no longer results in planets growing a lot more massive. This $t_{\textrm{sh}}$ threshold value is quite short for most of the embryos in a disc, meaning that  early, short initial shielding time is important in deciding the planet growth potential. Embryos  within 5.5\,au are able to accrete pebbles at earlier times compared to the outer embryos (as the pebble production front moves inside out), and hence have enough time to grow to high masses even without shielding. Similar trends can be observed for discs exposed to a wide radiation field ranging from F$_{\textrm{FUV,max}} = 10^5$ down to F$_{\textrm{FUV,max}} = 10^3$ G$_0$.

Compared to the single embryo per disc simulations where the embryo is assumed to be on a circular orbit throughout the simulation, in the multi-embryo per disc case the embryos start with randomised non-zero eccentricities and inclinations that evolve over time. As the eccentricity and inclination of a particular embryo affects the relative velocity of the pebbles approaching the embryo (see equation \ref{v_rel}), the pebble accretion rate onto this embryo also depends on the values of eccentricity and inclinations (see equation \ref{mdot_3D}  and \ref{mdot_2D}). Eccentric motions affect the initial accretion rates of the small embryos particularly strongly because they start by accreting pebbles in the Bondi regime, with the Bondi radius scales as $R_B \propto 1/v^{2}_{\textrm{rel}}$). This means some embryos with high initial pebble accretion rates can grow fast into proto-planets whilst others with low initial accretion rates stay embryo size throughout the simulations. Furthermore, compared to the single planet per disc scenario, some embryos in one disc can initially merge via collision, leading to more rapid growth. These embryos are then able to accrete at a faster rate and grow into proto-planets. All these additional stochastic effects of the N-body dynamics introduce random fluctuations in the total final masses made in the multi-planet systems in Fig. \ref{fig:total_m_final_made_selected_ainit_10^5G0}, compared to the smooth correlations between the final planet mass and either $t_{\textrm{sh}}$ or F$_{\textrm{FUV, max}}$ seen in the single planet simulations \citep[e.g. figure 5 in][]{2023MNRAS.522.1939Q}. Even so, the overall impacts of the reduced planet growth due to external photo-evaporation and the initial shielding time in facilitating planet growth in strongly irradiated discs are still clear, for discs with the fiducial initial disc mass of M$_d$ = 0.1\,M$_{*}$.

\begin{figure*}
    \centering
    \includegraphics[width=2\columnwidth]{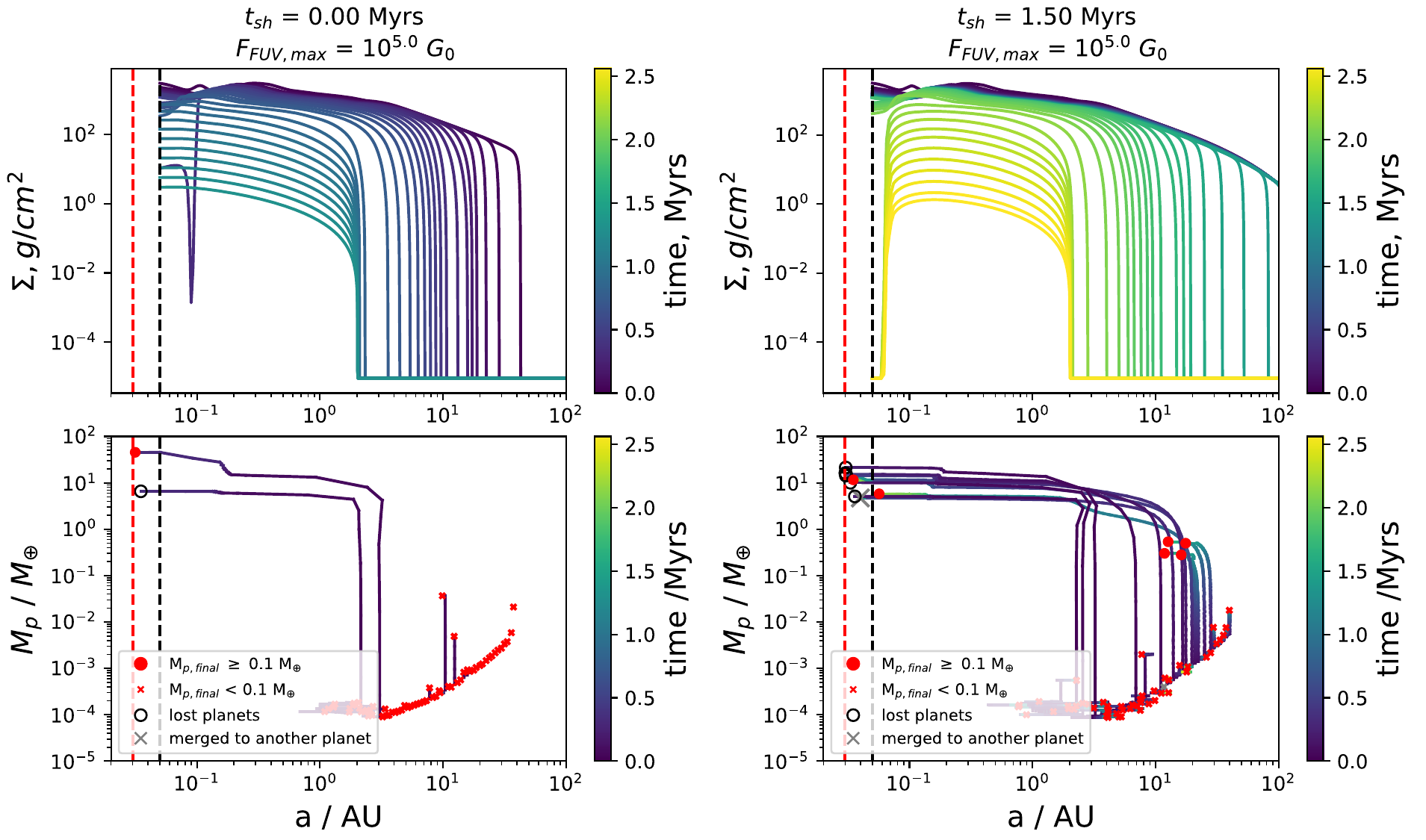}
    \caption{Top panels: evolution of surface density (with lines plotted every 50000 years) for discs with initial mass $M_{d}$ = 0.1\,M$_{\oplus}$ which are subject to an external FUV radiation field strength of F$_{\textrm{FUV,max}} = 10^5\,G_0$. The left panel shows the disc that is not shielding and the right panel shows the disc that is shielding for 1.5\,Myr. The bottom panels show the evolution of mass $M_p$ and semi-major axis of all the planetary embryos in the disc. The colour indicates the simulation time, and only the first 2.5 million years of the evolution is plotted, as the rapid disc depletion by the strong FUV field makes the gas disc evolution cease after only 1.3 Myr for the left hand non-shielded disc, and only 2.56 Myr for the right hand disc with $t_{\textrm{sh}}$ = 1.5 Myr. The red dashed line indicates the inner grid domain edge, and grey dashed line indicates the inner disc edge. On the bottom panels, the red solid circle and black empty circle indicate bodies that grew to a final mass $\geq 0.1$\,M$_{\oplus}$. The red solid circle indicates the final mass and semi-major axis of the ones which survived in the disc by the end of the simulation, the black empty circle indicates the final mass and semi-major axis of the ones that were lost during the simulation due to planet-planet interactions. The small red cross indicates the final mass and semi-major axis the bodies with final mass $< 0.1$\,$M_{\oplus}$, and the big grey cross indicates the bodies that merged with another body via collision. } 
    \label{fig:MvAandSigma_evol_combined}
\end{figure*}

Figure \ref{fig:MvAandSigma_evol_combined} illustrates the evolution of the disc surface density (top panels), along with the evolution of mass and semi-major axis of planets (bottom panels) in two strongly irradiated discs (F$_{\textrm{FUV,max}} = 10^5$\,G$_0$), one without shielding (left-hand panels) and the other with a shielding time of 1.5\,Myr (right-hand panels). The disc without shielding (left-hand panels) highlights the effect of external photo-evaporation in curbing the growth of planets: the top panel demonstrates the rapid depletion of the outer disc edge down to smaller than 10\,au in just 0.1\,Myr without shielding, starving the embryos especially those at larger semi-major axis of the available pebble reservoir for growth. In contrast, more embryos in the shielded disc in the right-hand panels of Fig. \ref{fig:MvAandSigma_evol_combined} are able to grow to planet size and migrate especially for the embryos initially located in the outer disc region, as the fast radius truncation is delayed until after cloud shielding. The bottom plots of Fig. \ref{fig:MvAandSigma_evol_combined} also show the stochastic nature of planet formation in the multi-planet per disc system: regardless of the shielding time, only some of the initial embryos with faster initial pebble accretion rates due to eccentricity and initial collisional merging are able to grow to planet size, while a lot more embryos with low initial accretion rates did not get the chance to grow significantly throughout the simulation. 

The effects of external photo-evaporation on limiting planet growth potential are expected to vary depending on the initial disc mass, which specifies the amount of planet-forming material available in the disc. Discs with larger initial masses with larger pebble reservoirs can facilitate a larger number of embryos growing into massive planets and be more resilient to the effects of external photo-evaporation. On the other hand, discs with small initial masses have a limited supply of pebbles available for embryos to accrete from even before radius truncation, so the effects of external photo-evaporation might not be significant.

\begin{figure*}
    \centering
    \includegraphics[width=2\columnwidth]{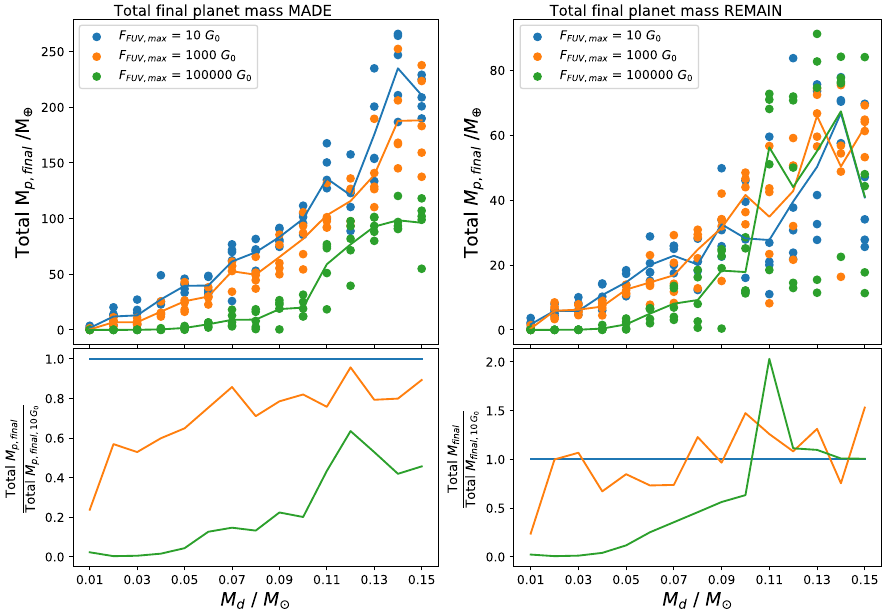}
    \caption{Plots of final planet masses in discs of varying initial masses that are subject to different FUV radiation strengths indicated by different colors: 10\,G$_0$ (blue), $10^3$\,G$_0$ (orange), $10^5$\,G$_0$ (green). Top panels: the top left panel shows the total final planet masses formed in discs regardless of whether the formed planets were later lost; the top right panel shows the total final planet masses left in discs at the end of simulations after some planets were lost due to planet-planet interactions. The dots shows the value of total final planet mass in each of 5 runs with the same parameters, and solid line shows the averaged value of the 5 runs. Bottom panels: bottom left panel shows the ratio of final planet mass made (averaged over 5 runs) in discs subject to different radiation strengths compared to the planets made in disc subject to weakest radiation (10\,G$_0$); bottom right panel shows the ratio of the final planet mass remaining (averaged over 5 runs) in discs subject to different radiation strengths compared to the planets remaining in discs subject to the weakest radiation (10\,G$_0$).}
    \label{total_mp_varMdisc_set3}
\end{figure*}

The left-hand panels of Fig. \ref{total_mp_varMdisc_set3} show the effects of external photo-evaporation on limiting planet growth for discs with varying initial disc masses. The top left-hand panel in Fig. \ref{total_mp_varMdisc_set3} plots the total final planet masses M$_{\textrm{p, final}}$ that all the embryos in a disc grew to, as a function of the initial disc mass M$_{d}$. Different colours indicate discs that are subject to different levels of FUV radiation flux. The dots indicate the results of each of the 5 realisations with the same set of parameter values, and the solid lines represent the averaged value over the 5 realisations. The plot indicates that the planet growth potential via pebble accretion is sensitive to both initial disc mass and external photo-evaporation. Across all UV field strengths, the total planet mass achieved by embryos increases with the initial disc mass as expected, and for all initial M$_d$, stronger UV radiation results in lower total planet masses. However, the extent to which external photo-evaporation limits planet growth varies with initial disc mass. This is most noticeable for discs irradiated with the strongest UV field of F$_{\textrm{FUV,max}} = 10^5$\,G$_0$ (the green curve). In this case the rapid radius truncation due to high external photo-evaporative mass loss rate suppresses planet growth very effectively for discs with initial M$_d$ of up to 0.1\,M$_\odot$, limiting the total M$_{\textrm{p, final}}$ made per disc under just $\sim$ 10 M$_\oplus$. However, discs with initial M$_{d} > 0.1$\,M$_{\oplus}$ are still able to facilitate the formation of planets that are massive enough to preserve the clear correlation of higher final total $M_{\textrm{p,final}}$ made in discs with higher M$_d$. The varying degrees of effectiveness of external photo-evaporation on preventing planet growth as a function of initial M$_d$ is more clearly highlighted in the bottom left-hand panel of Fig. \ref{total_mp_varMdisc_set3}, which plots the the total $M_{\textrm{p, final}}$ of planets that form in discs subject to different $F_{\textrm{FUV, max}}$, as a ratio to those that formed in the least irradiated discs ($F_{\textrm{FUV, max}}$ = 10\,G$_0$). Compared to planets formed in discs with weak radiation (10 G$_0$), stronger FUV radiation prohibits planet growth much more effectively in less massive discs. In the case of F$_{\textrm{FUV, max}} = 10^5$\,G$_0$, in discs with initial $M_d$ < 0.1\,M$_{\odot}$, the total $M_{\textrm{p, final}}$ formed were only $<20$ \% of the masses formed in the weakly irradiated discs (10\,G$_0$). But in discs of initial M$_d$ > 0.11\,M$_{\odot}$, more than 40 \% of the total M$_{\textrm{p, final}}$ were still able to form compared to the weakly irradiated cases. A similar trend can be observed for discs exposed to the medium UV radiation field of $10^3$\,G$_0$. In summary, Fig. \ref{total_mp_varMdisc_set3} shows that planet growth via pebble accretion depends on both the initial disc mass and FUV field strengths, but strong external photo-evaporation becomes the dominating factor in suppressing planet growth in less massive discs.

\subsection{The effects of planet-planet interactions and planetary evolution}
\label{plane_interactions}
\begin{figure}
    \centering   
    \includegraphics[width=\columnwidth]{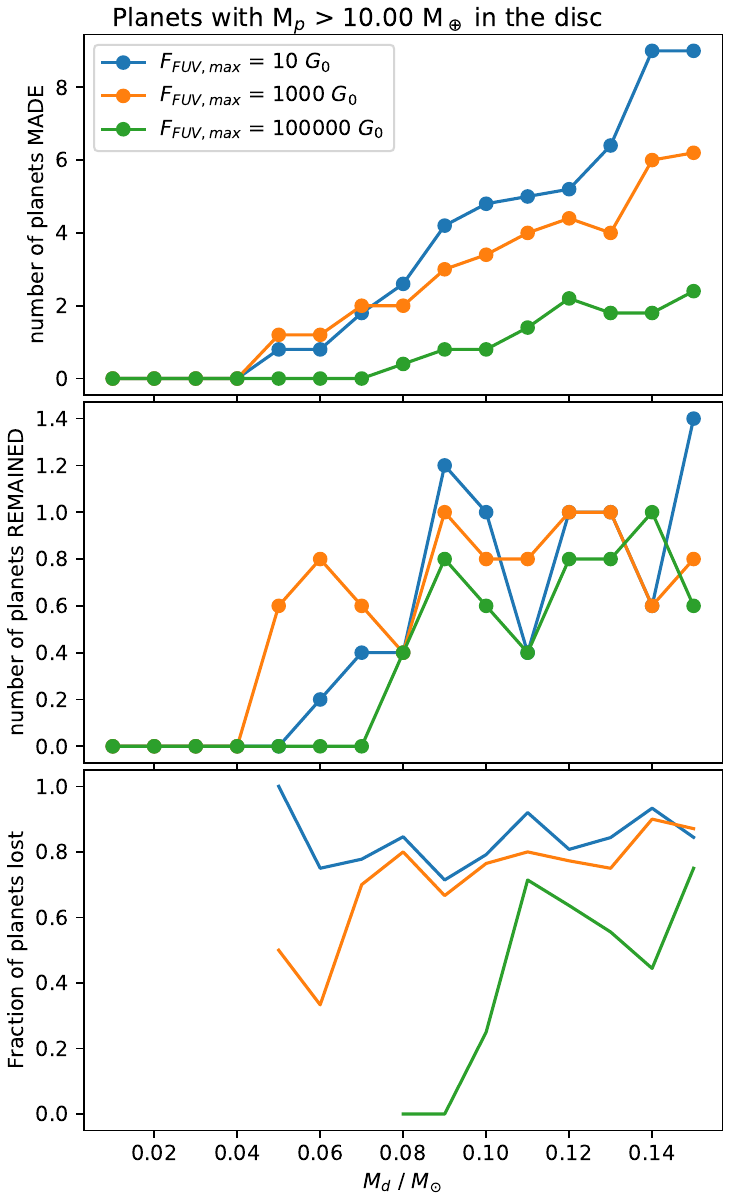}
    \caption{Number of planets (mass $M_p \geq$ 10\,M$_\oplus$) per disc (averaged over 5 realisations of the same parameters) as a function of varying initial disc masses, radiated by different FUV radiation strengths indicated by different colors: 10\,G$_0$ (blue), $10^3$\,G$_0$ (orange), $10^5$\,G$_0$ (green). Top panel: number of planets with M$_p \geq$ 10\,M$_\oplus$ formed in discs, regardless of whether the formed planets were later lost. Middle panel: number of planets with $M_p \geq$ 10\,M$_\oplus$ left in discs with varying masses at the end of simulations after some planets lost due to planet-planet interactions. Bottom panel: fraction of the formed planets that were lost due to planet-planet interaction in discs with varying initial masses.}
    \label{number_planets_set3_above10Mearth.pdf}
\end{figure}

The previous section focused on analysing how external photo-evaporation acts to suppress planet growth potential via pebble accretion in discs containing multiple embryos. The simulations showed that planet growth and migration are sensitive to the external radiation field strengths and shielding times across a wide range of disc masses, but external photo-evaporation is particularly effective in restraining planet growth for discs with smaller initial mass. However, the formation and dynamical evolution of the multi-planet systems are a combined result of pebble accretion, gas accretion and planet-planet interactions. Some of the planetary embryos that were able to grow to planet sizes can later be lost as a result of dynamical scattering and ejections, as well as forced migration past the inner disc boundary on to the central star. Planets formed in the disc which can survive until the end, and hence the final resultant architecture, are therefore an outcome of both early stage external photo-evaporation and cloud shielding, and the later stage dynamical evolution of formed planets. 

In the N-body planet formation model used in this paper, the main way of losing planets in a disc is via migration onto the central star, or through planet--planet collisions. For a planet that grows massive enough and migrates inward, it will eventually be halted at the disc inner edge (0.05\,au) or in the magnetospheric cavity (see section \ref{mag_cavity} for details). However, a second inwardly migrating planet coming close to the first halted planet can push the planet further inward to a shorter period orbit via resonant gravitational interactions. If the first planet got nudged to a semi-major axis that is smaller than the grid domain (0.03\,au), it is assumed to have migrated on to the star. This effect can be seen in the two disc examples in Fig. \ref{fig:MvAandSigma_evol_combined}. The bottom panels show the evolution of mass and semi-major axis of planets in a disc of initial mass $0.1$ M$_\odot$ irradiated by a F$_{\textrm{FUV,max}} = 10^5$\,G$_0$ field. The disc in the right-hand panel in Fig. \ref{fig:MvAandSigma_evol_combined}, which was shielded for 1.5\,Myr, facilitated the formation of more massive (> 10\,M$_{\oplus}$) planets which migrated inwards and halted in the magnetospheric cavity (cavity edge shown by the black dashed line), compared to the non-shielded disc in the left hand panel of Fig. \ref{fig:MvAandSigma_evol_combined} (as mentioned in section \ref{planet_forming_potential}). However, most of these halted planets in the right hand shielded disc were later lost by being pushed beyond the inner disc domain (shown by the red dashed line) by another inwardly approaching planet, so the final planets that remained in the shielded disc ended up being similar to those in the weakly irradiated disc (left-hand panel of Fig. \ref{fig:MvAandSigma_evol_combined}).
In a disc, if a larger number of planetary embryos are able to grow more massive due to higher initial disc mass or lower FUV radiation strength (as discussed in the previous section), there would be more frequent gravitational interactions among the planets, resulting in more planets lost due to such interactions. This effect is illustrated in Fig. \ref{number_planets_set3_above10Mearth.pdf} where the top panel plots the number of planets with mass M$_p \geq$ 10\,M$_\oplus$ formed per disc (averaged over 5 realisations of the same parameters), regardless of whether they are later lost, as a function of initial disc mass, clearly showing the trend of more massive (M$_p \geq$ 10\,M$_\oplus$) planets formed in more initially massive, less irradiated discs. However, the bottom panel, which plots the fraction of formed planets that were later lost, shows that in discs where an average of more than 3 massive ($\geq 10$\,M$_\oplus$) planets were formed, more than 40\% were lost (for all values of F$_{\textrm{FUV, max}}$). Overall, there are higher fractions of planets lost in more initially massive, less irradiated discs. For discs in the weakest FUV field of 10\,G$_0$ for example, at least 60 \% of the massive (> 10\,M$_\oplus$) planets formed in discs of initial mass > 0.08 M$_{\odot}$ were lost. As a result, the middle panel of Fig. \ref{number_planets_set3_above10Mearth.pdf}. which plots the number of planets with $M_p \geq$ 10\,M$_\oplus$ remained in disc at the end (excluding the ones lost in the evolution process), shows that for discs with initial mass > 0.08\,M$_\odot$, the number of survived planets no longer correlates with either initial disc mass or FUV field. 

Similar trends can be observed in terms of the total survived M$_{\textrm{p, final}}$, as a function of the initial disc mass M$_d$, as plotted in the top right-hand panel of Fig. \ref{total_mp_varMdisc_set3}. Different colours indicate different external FUV field strengths. (Recall that the left-hand panels include all bodies, including those lost to the star.) Although an overall trend of more massive discs retaining a higher total survived $M_{\textrm{p, final}}$ is still visible, the survived total $M_{\textrm{p, final}}$ in discs with initial $M_{d} \gtrsim$ 0.08\,M$_{\odot}$ becomes much more random with no visible correlation between $F_{\textrm{FUV, max}}$ and the total $M_{\textrm{p, final}}$ of planets remaining in the initially more massive ($\gtrsim 0.08\,M_{\oplus}$) discs. However, for discs with initial M$_{d} < 0.08$\,$M_{\odot}$, the effects of stronger external photo-evaporation suppressing the surviving planet mass is still visible as highlighted in the bottom right-hand panel of Fig. \ref{total_mp_varMdisc_set3}, which plots the ratio of total M$_{\textrm{p, total}}$ of planets remaining in discs irradiated by different $F_{\textrm{FUV, max}}$, compared to the discs subject to the weakest radiation of 10\,$\rm G_0$. Discs of initial mass < 0.08\,$M_{\odot}$ that are irradiated by the strongest field of $10^5$ G$_0$ (in green) contained less than half of the total survived planet mass, compared to their weakly irradiated counterparts, as a result of fewer planets being able to form in strongly photo-evaporated discs in the first place. In the medium FUV field strength of $F_{\textrm{FUV, max}} = 10^3 \rm G_{0}$, discs of the low initial mass range (0.03--0.08\,$M_{\odot}$) also see at least 20\% lower total surviving planet mass compared to the same discs that are weakly irradiated. Overall, the final resultant total mass of multi-planet systems that form via pebble accretion in externally photo-evaporating discs is dominated by planet-planet interactions in more massive discs, while the effects of external photo-evaporation can still be visible in reducing the total $M_{\textrm{p, final}}$ retained in the less massive discs.  {However, as described in section \ref{mag_cavity}, we simulated the effect of a planet trap at disc inner edge by removing all migration torques on a planet once it reaches the inner edge (such that the planet halts at this location). It is worth noting that our choice of the surface density profile and an open inner boundary condition does not produce the positive migration torque at inner edge to resist the halted planet being pushed into cavity by the resonance pushing of other planets (especially when multiple planets migrate inwards together as part of a resonant chain). In some hydrodynamic models of disc \citep[e.g.][]{2023MNRAS.523.3569Y, 2025arXiv251001332H}, the choices of surface density and viscosity transition at disc inner edge regions yield strong enough positive migration torque so that it can resist resonance pushing of planets and halt multiple planets in resonant chains. Therefore, depending on the detailed modelling of the inner edge region, more planets can be retained by the planet trap at the inner edge and survive, rather than being lost onto the central star. In such cases, the effects of external photo-evaporation in reducing the total M$_{\textrm{p, total}}$ in massive discs might also be visible when weakly irradiated discs can retain more planets due to the planet traps at their inner edges.}

\section{Resulting Planetary Architectures}

The previous section focused on analysing the impacts of external photo-evaporation on the formation and evolution of planet masses in irradiated discs. This section analyses the impacts of external photo-evaporation on the resultant planetary architectures from the simulations.

\subsection{Close in (< 1 au) planet population}
\label{close_in_planet}
\begin{figure*}
    \centering
    \includegraphics[width=1.35\columnwidth]{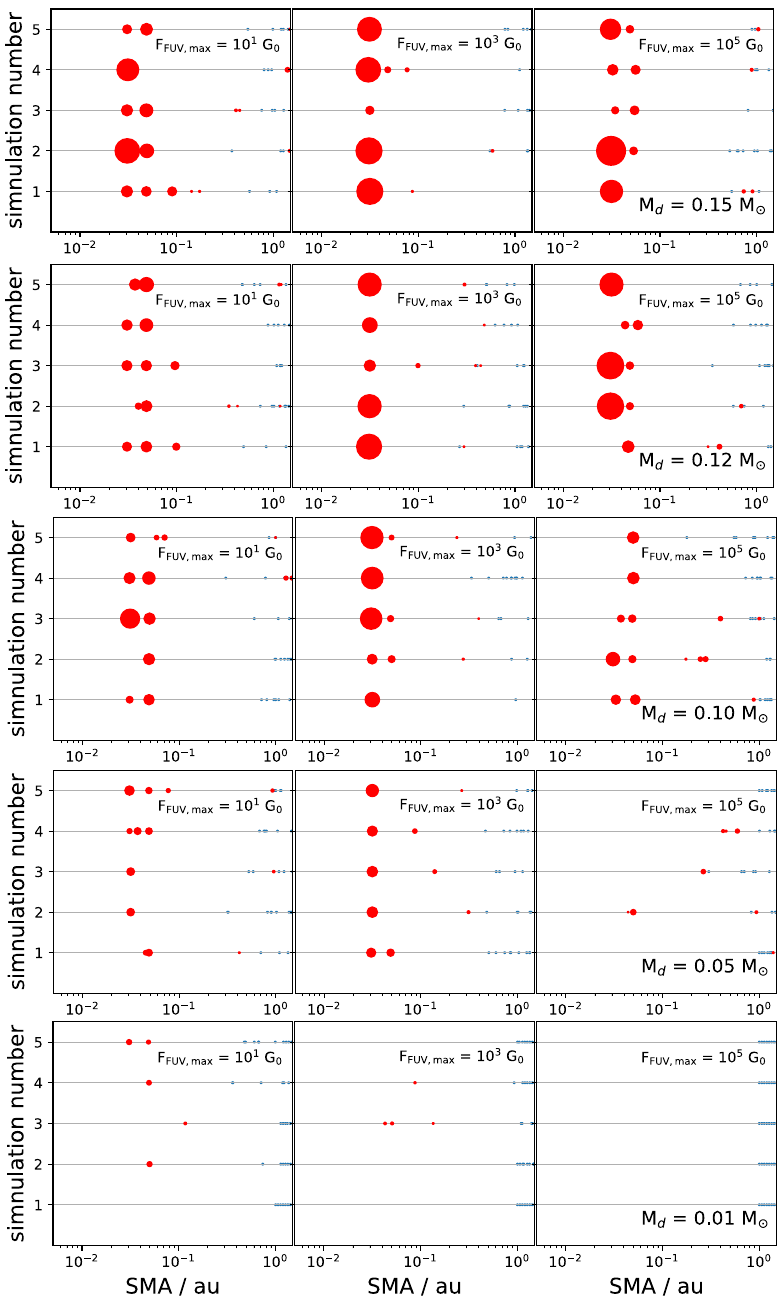}
    \caption{Planets formed in the inner 1\,au of discs of varying disc mass and external FUV field strengths. Each row represents discs of a particular initial disc mass, each column represent a particular value of the external FUV field $F_{\textrm{FUV, max}}$. Each panel contains all 5 realisations (labeled as simulation number 1 - 5 on the vertical axis) of a given disc mass and $F_{\textrm{FUV, max}}$. The horizontal axis represents the final semi-major axis location of the planets, and the circle size is proportional to the final mass of the planet. The red solid circles represent the planets (with mass $\geq$ 0.1\,M$_{\oplus}$) that survived in the disc at the end of the simulation. The small blue dots represents bodies that remained planetary embryos with mass < 0.1\,M$_{\oplus}$. }
    \label{planet_table_set3.pdf}
\end{figure*}

Figure \ref{planet_table_set3.pdf} shows examples of the final planet populations formed in the inner 1 au of discs with varying initial disc masses (rows) and external FUV radiation strengths (columns). The inner 1 au of the discs contain most of the massive planets formed in the simulations due to inward migration. In each row (of a certain disc mass), the three panels from left to right show planets formed in the radiation environment of F$_{\textrm{FUV, max}} = 10$ G$_0$ (left), $10^3$ G$_0$ (middle) and $10^5$ G$_0$ (right). Each panel contains the 5 realisations (labelled as simulation number 1 - 5 on the vertical axis) of a given initial disc mass and given F$_{\textrm{FUV, max}}$. In each panel, final planet mass is represented by the circle size,  and the final semi-major axis is represented by the horizontal axis. 

The top three rows show the population survived in the discs of three larger initial masses (0.15, 0.12 and 0.1 M$_\odot$). In these discs, there is no discernible difference in the remaining planet population from left to right across the weak to strong FUV fields. On the other hand, in the bottom two rows showing the cases of less massive discs (0.05 and 0.01 M$_\odot$), the trend of decreasing number of planets and decreasing planet mass can be seen from left to right, with stronger FUV field. As discussed in the previous section, more massive discs retain pebble reservoirs for longer that facilitate the formation of more planets of higher masses (most of which then migrate inwards to the inner regions of the disc during the simulation). This results in more frequent gravitational interactions among the planets as they congregate at migration traps such as the inner disc edge. In these initially more massive discs, even though more massive planets were able to form in discs with weaker external photo-evaporation, the increased frequencies of gravitational encounters resulted in more planets being lost via migration on to the star due to being pushed by other inwardly migrating planets. The resultant planetary architectures of the survived planets in the inner au in weakly irradiated discs become indistinguishable to those that survived in strongly irradiated discs, where fewer massive planets formed. On the other hand, in less massive discs, which have smaller initial pebble mass reservoirs, generally fewer massive planets formed in comparison, meaning less frequent planetary interactions. Therefore the planets that formed are less prone to being lost in weakly irradiated discs, such that the impact of external photo-evaporation can still be visible when the rapid radius truncation results in even fewer planets forming and migrating to the inner au in strongly irradiated discs.  {However, as mentioned in section \ref{plane_interactions} the impact of external photo-evaporation on the close in planet population in more massive discs might be more visible if positive migration torques at inner-disc edges are strong enough to retain more planets. However, in this case the final close planetary architecture would still be dominated by gravitational interactions. especially with more survived planets in inner-disc regions.}



\subsection{Wide orbit (> 10au) planet population}
\label{wide_planet}
The population of close orbit planets consists mainly of planets formed by embryos initially located within 10\,au that have grown massive enough ($\gtrsim $ 1\,M$_\oplus$) to inwardly migrate to the region. The inner au region planets therefore are more affected by gravitational interactions among each other. In contrast, the planets that ended up in the wide orbit region of > 10 au are resulted from the embryos initially located at > 10 au that grew to only $\sim $ 0.1--1\,M$_\oplus$. In the pebble accretion mechanism, a planetary embryo starts to receive an influx of pebbles to grow once the pebble production front, which propagates inside-out, reaches its initial semi-major axis location. The embryos at larger orbits therefore start to receive pebble flux later compared to the inner ones and therefore only have enough time to grow to smaller terrestrial sizes, even in an isolated disc without external photo-evaporation. Compared to the close-orbit region with more planets that migrated inward, there are fewer planets populating the wide-orbit region, such that they encounter less frequent gravitational interactions, making the impact by external photo-evaporation more prominent regardless of the initial disc mass.

\begin{figure*}
    \centering
    \includegraphics[width=1.5\columnwidth]{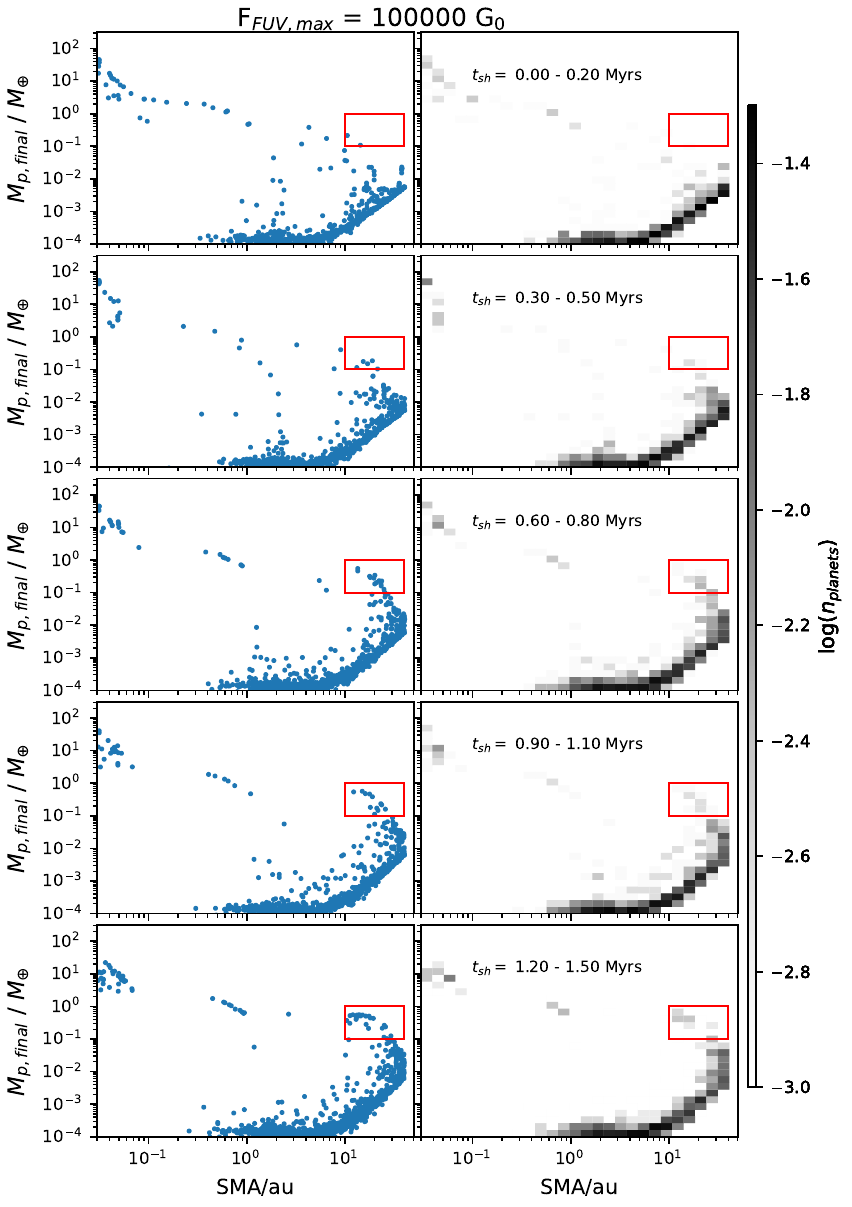}
    \caption{Left columns: Scattered plot of final mass and semi-major axis of all bodies (planets and embryos) remaining at the end of the simulation in the discs irradiated by $F_{\textrm{FUV, max}}$ = $10^5$\,G$_0$ with different ranges of shielding time $t_{\textrm{sh}}$ from top to bottom. Right columns: The number density of resulting bodies as a function of final mass and semi-major axis in discs irradiated by $F_{\textrm{FUV, max}}$ = $10^5$\,G$_0$ with different ranges of $t_{\textrm{sh}}$ from top to bottom.}
    \label{fig:Heat_map_10^5G0}
\end{figure*}

The impact of external photo-evaporation and shielding time on the final planetary population in the wide orbit (> 10\,au) region is shown in Fig. \ref{fig:Heat_map_10^5G0} that shows the resultant planet populations from the set of simulations with the fiducial initial disc mass of 0.1\,M$_{\odot}$ with FUV field strength of $10^5 \,\rm G_0$. The left column panels are scatter plots of the final mass and semi-major axis of all bodies (planets and embryos) remaining at the end of the simulation with increasing shielding time $t_{\textrm{sh}}$ from top to bottom. The right-hand panels show the number density heatmap of all bodies included in the corresponding left panels, with darker shades indicating more bodies populated in that region of final mass and semi-major axis. It can be seen that the region of SMA$_{\textrm{final}} > 10$\,au and $0.1 < M_{\textrm{p, final}} < 1$\,M$_{\oplus}$ (i.e. wide orbit small terrestrial planets highlighted by the red rectangles) becomes more populated as the shielding time increases from top to bottom. As mentioned earlier, the time available for an embryo in a disc to accrete pebbles is the time between the pebble production front reaching its initial semi-major axis and moving past the disc outer edge (when the incoming pebble flux is cut off). If the disc is then subject to fast outer edge truncation by external photo-evaporation, the pebble production front will move out of the truncated disc edge even quicker, shortening the pebble accretion time for the embryo. Since the embryos on more distant orbits start receiving pebble flux much later than those closer to the star, how much they get to grow is much more sensitive to the pebble cutoff time as a result of the radius truncation. Thus in discs that are shielded for longer, by delaying the pebble cutoff time, more of these embryos can grow to planet size, populating the highlighted region in Fig. \ref{fig:Heat_map_10^5G0}. Similarly, in discs that are subject to a lower external mass loss rates in a lower FUV radiation field, the disc outer edge would truncate at a slower rate. Pebble accretion times for the outer embryos would therefore be longer to facilitate growth. 
 Similarly, in discs that are subject to a lower external mass loss rates induced by a weaker FUV radiation field, the disc outer edge would truncate at a slower rate. Pebble accretion times for the outer embryos would therefore be longer to facilitate their growth. See figure \ref{fig:Heat_map_threeGmax} for the comparisons of planet population resulted from different FUV field environments.

In summary, the impacts external photo-evaporation on the resultant planetary architecture of a multi-planet system can be seen in the inner au planet population from discs with small initial masses (< 0.1\,M$_\odot$). Stronger FUV field results in fewer less massive (1--10\,M$_\oplus$) planets remaining in this region. However, the inner au planet population from more initially massive discs ($\geqslant$ 0.1\,M$_\odot$) is dominated by planet-planet interactions, masking any visible impacts of external photo-evaporation. The impacts of external photo-evaporation can also be seen in the wide orbit (> 10 au) region, with more small terrestrial planets (0.1--1\,M$_{\oplus}$) populating this region in a disc.

\subsection{The impact of external photo-evaporation on resonance occurrence rate}
\begin{figure}
    \centering	\includegraphics[width=\columnwidth]{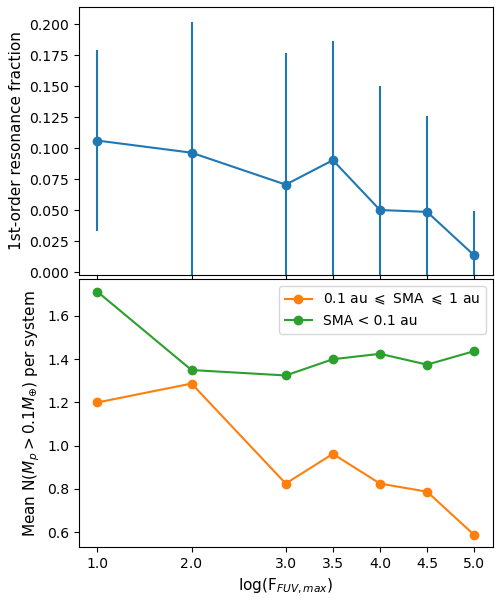}
    \caption{top panel: each data point is the mean value of fraction of 1st-order resonant pairs as a function of from all simulation runs with the corresponding F$_{\textrm{FUV,max}}$ on the x-axis, with the error bar showing the standard deviation. Bottom panel: mean number of planets survived per planetary system with final planet mass M$_p$ > 0.1\,M$_\oplus$, the green curve shows the ones with final semi-major axis < 0.1\,au, these planets typically have planet mass between 1 - 100\,M$_{\oplus}$, the orange curve shows the ones with final semi-major axis within between 0.1 and 1\,au these planets typically have planet mass between 0.1 to 10\,M$_{\oplus}$. } 
    \label{fig:resonance}
\end{figure}

In this section we analyse the effects of external photo-evaporation on the occurrence of first-order resonance pairs in the resultant planetary systems formed. We identify first order resonant planet pairs as the adjacent planet pairs survived at the end of the simulation that firstly satisfy:  - 0.05 < $\Delta$ < 0.05, with $\Delta$ being the fractional deviation from perfect period commensurability 
    \begin{equation}
    \Delta \equiv \frac{P_{\textrm{out}}/ P_{\textrm{in}}}{p/q} - 1
    \end{equation} where $P_{\textrm{in}}$ and $P_{\textrm{out}}$ are the orbital periods of the inner and outer adjacent planet pairs in the simulation, $p$ and $q$ and are small positive integers that define a given MMR \citep[see e.g.][]{2010MNRAS.405..573P}, and secondly with resonance amplitude < $180^{\circ}$.

In all the planetary systems simulated, generally there are two populations of first-order resonance pairs. The first population is close-in more massive planets (between 1 to 100\,M$_\oplus$) in the inner 0.1\,au, and the second population is between $0.1 - 1$\,au, typically with final planet mass typically between 0.1 to 10\,M$_\oplus$. As shown in Sect. \ref{close_in_planet}, planets lost due to gravitational interactions resulted in similar numbers of surviving close-in massive planets across all values of F$_{\textrm{FUV, max}}$, this trend can be seen in the green curve in the bottom panel of Fig. \ref{fig:resonance}, which plots the mean number of planets per system with final semi-major axis within 0.1\,au. As a result, the number of resonant pairs in the first population stays similar across all ranges of F$_{\textrm{FUV,max}}$. However, the survived planets remaining between 0.1 and 1\,au still decreases with higher F$_{\textrm{FUV,max}}$, reducing the number of available planets getting into resonant pairs in this location range, which is reflected in the orange curve plotting the mean number of planets per system with final semi-major axis between 0.1 and 1\,au. From this decrease we therefore expected a decrease in first order resonance fraction as a function of increasing F$_{\textrm{FUV, max}}$. The top panel of Fig. \ref{fig:resonance} plots the mean value of first-order resonance fraction as a function of FUV field strengths from the simulations of set 1 with fixed disc mass of 0.1\,M$_*$. Even though an overall trend of decreasing resonance fraction with stronger FUV field strengths can be seen, which traces the shape of the orange curve in the bottom plot, each mean value at a specific F$_{\textrm{FUV,max}}$ has a quite large standard deviation value as denoted by the error bar.  {The large uncertainties arise because first-order resonance pairs that survive until the end of the simulations are quite rare in all simulated systems regardless of the strength of external photo-evaporation. Many planets halted at the disc inner edge which enters resonance with another type I migrating planet got pushed onto the central star. Therefore given the large uncertainties the correlation between 1st-order resonance fraction and FUV field strength is very weak. However, this might be different if strong positive migration torques at disc inner edges are strong enough to trap and retain more planets entering resonant chains.}


\section{Summary and conclusions}
 
In this paper we investigate how the formation and evolution of multi-planet systems are affected by external photo-evaporation in a stellar cluster environment. We use a model of N-body simulations of multiple planet formation via pebble accretion coupled with a 1-D viscous disc subject to external photo-evaporation. Compared to the single planet per disc model used generally in the previous research, the effects of gravitational interaction between planets are included. With this model we investigate the impacts of external photo-evaporation and cloud shielding on the planet growth via pebble accretion as well as the resulting planetary architecture resulted in star-disc systems in a clustered birth environment. We draw the following main conclusions from this study:

\begin{enumerate}
\item As expected from the results of the single planet per disc simulations from \citet{2023MNRAS.522.1939Q}, external photo-evaporation still acts to reduce the planet growth in mass via reducing the pebble mass reservoir in discs containing multiple planetary embryos across a wide range of disc masses, and is particularly effective in restraining planet growth in lower mass discs (< 0.1\,M$_{\odot}$). A short initial shielding time ($\sim 0.3$\,Myr) is
effective in facilitating mass growth of multiple planetary system via pebble accretion especially for the less massive discs. 

\item In terms of the final total mass remaining in each system after dynamical evolution, planets lost due to planet-planet interactions dominate the outcome for planets in more massive ($\geq$ 0.1\,M$_{\odot}$) discs, masking the effects of external photo-evaporation. However, the effects of external photo-evaporation is still visible in reducing the total final mass retained in the less massive(< 0.1\,M$_{\oplus}$) discs.  {However, such trend might also be observed in more massive discs if more planets can be retained by the positive migration torques at inner-disc edges.}

\item In terms of the final resulting planetary architectures, for the resulting planet populations surviving in the inner 1\,au, there is no statistically significant signature of external photo-evaporation in initially more massive ($\geq$ 0.1\,M$_{\odot}$) discs, 
because of the planets lost to the central star prompted by interactions with another approaching planet.  {Again, this might be different if more planets can be retained by the positive migration torques at inner-disc edges, however the inner au planet architecture will still likely to be dominated by the outcome of later stage planet-planet interactions.} In less massive (< 0.1\,M$_{\odot}$) discs, the signature of external photo-evaporation is visible, with planets formed and surviving in discs irradiated with stronger FUV field being less massive, and fewer in number. External photo-evaporation also leaves a signature for the wide orbit (> 10 au) terrestrial planets (0.1 - 1\,M$_{\oplus}$), with less planets populating this region for stronger FUV field/shorter shielding time. 

\item  {or 1st-order resonant pairs fraction, large uncertainties arise as most resonance pairs are lost onto the central star and rarely survive until the end of the simulations for all values of FUV field strength. Consequently a clear correlation between 1st-order resonance fraction and FUV field strength can not be claimed.}
\end{enumerate}


In summary, we demonstrated that external photo-evaporation limits the growth of multi-planet systems via pebble accretion in stellar clusters, but the evolution and outcome of the resultant planetary architectures depends on both the early stage external photo-evaporation of the disc and the later stage dynamical evolution of planets. The impacts of external photo-evaporation on the resultant planet population are more visible if they are formed in less initially massive (< 0.1\,M$_\odot$) discs. However, if the disc inner edge is able to retain more planets, then the correlation between more remaining planets with weaker external photo-evaporation might also be visible in more massive discs. Here we used a parameterised time-varying FUV track with a constant field strength ($F_{\textrm{FUV,max}}$) after a shielding period to simulate the radiation environment of a stellar cluster, in a dynamically evolving star forming region, the FUV field radiated upon a disc constantly varies with time \citep{2022MNRAS.512.3788Q, 2023arXiv230203721W}. In order for a realistic study of to what extent the formation and evolution of multi-planet systems are affected by the cluster they are born in, planet formation models should be coupled to the FUV tracks traced from stellar cluster and feedback simulations. In this study, for each set of parameters, we only ran 5 realisations with randomised initial 3D planet positions and velocities due to computational cost. While our simulations have qualitatively demonstrated the impacts of external photo-evaporation, it is not enough to quantify the impacts on terrestrial planet population between the observable range between 1-10\,au. Future research can double the number of runs for each parameter sets, in order to statistically quantify the impacts of FUV field strengths on the resultant planet population especially in the observable range of within 10\,au.

\section*{Acknowledgements}
LQ and TJH acknowledge UKRI guaranteed funding for a Horizon Europe ERC consolidator grant (EP/Y024710/1). TJH also acknowledges funding from a Royal Society Dorothy Hodgkin Fellowship.  
GALC was funded by the Leverhulme Trust through grant RPG-2018-418.
This work was performed using the DiRAC Data Intensive service at Leicester, operated by the University of Leicester
IT Services, which forms part of the STFC DiRAC HPC Facility (www.dirac.ac.uk). The equipment was funded by BEIS capital
funding via STFC capital grants ST/K000373/1 and ST/R002363/1
and STFC DiRAC Operations grant ST/R001014/1. DiRAC is part
of the National e-Infrastructure.
This research utilised Queen Mary's Apocrita HPC facility, supported by QMUL Research-IT (http://doi.org/10.5281/zenodo.438045).

\section*{Data Availability}
The simulation results are available upon request to the corresponding author.



\bibliographystyle{mnras}
\bibliography{Planet_form_paper} 




\appendix

\section{Plant population figures for varying FUV field strengths}
\begin{figure*}
    \centering
    \includegraphics[width=2.1\columnwidth]{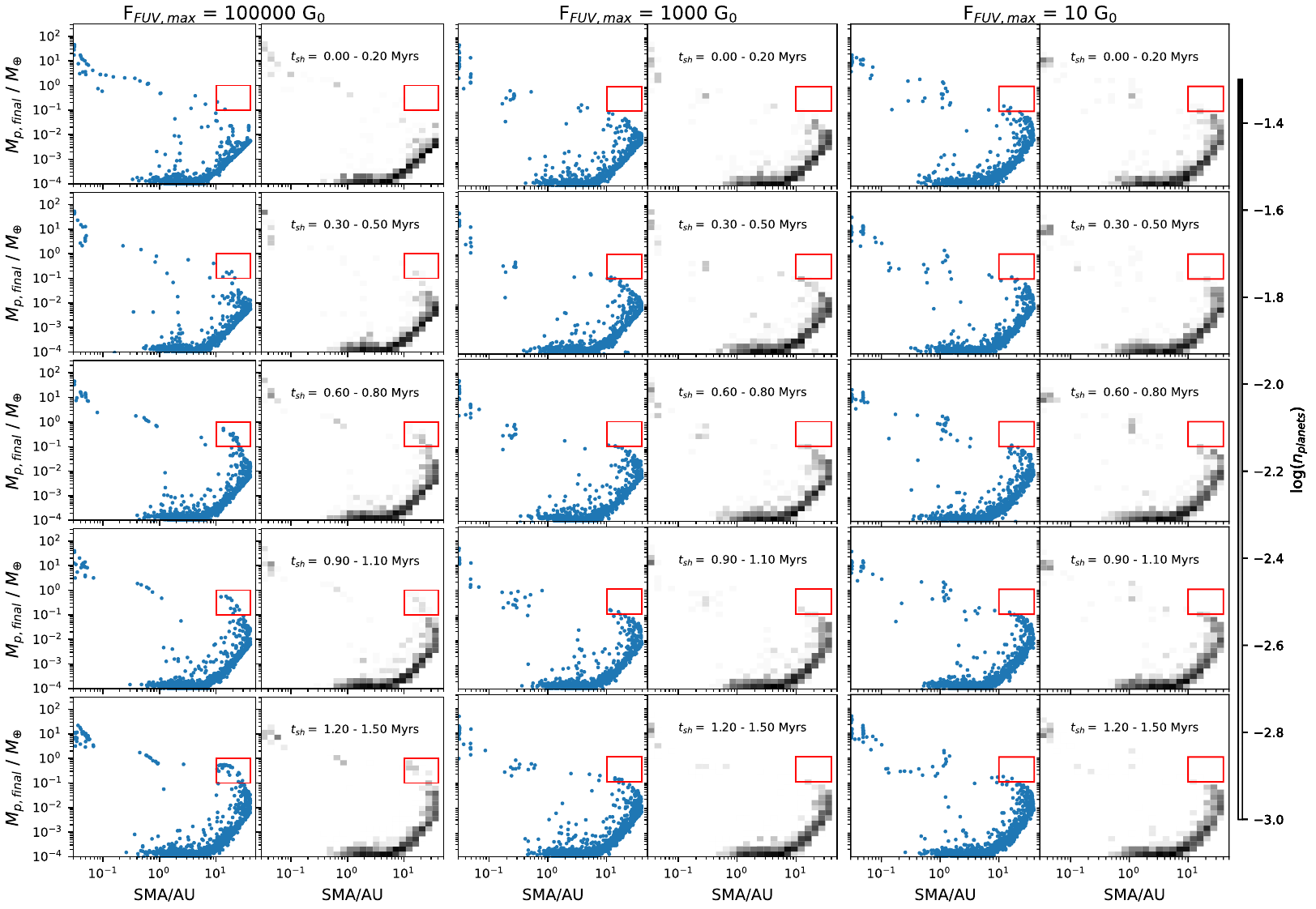}
    \caption{Left side columns of each of the three subplots: Scattered plot of final mass and semi-major axis all bodies (planets and embryos) remaining at the end of the simulation in the discs irradiated by $F_{\textrm{FUV, max}}$ = $10^5$ G$_0$ (the left subplot), $10^3$ G$_0$ (the middle subplot) and 10 G$_0$ (the right subplot) with different ranges of shielding time $t_{\textrm{sh}}$ from top to bottom. Right columns of each of the three subplots: the number density of resulting bodies as a function of final mass and semi-major axis in discs irradiated by $F_{\textrm{FUV, max}}$ = $10^5$ G$_0$ (the left subplot), $10^3$ G$_0$ (the middle subplot) and 10 G$_0$ (the right subplot) with different ranges of $t_{\textrm{sh}}$ from top to bottom.}
    \label{fig:Heat_map_threeGmax}
\end{figure*}

Similar to Fig. \ref{fig:Heat_map_10^5G0}, Fig. \ref{fig:Heat_map_threeGmax} shows the resultant planet populations from the set of simulations with the fiducial initial disc mass of 0.1\,M$_{\odot}$ for a range of FUV field strengths: $10^5 \rm G_0$ (columns 1 and 2), $10^3 \rm G_0$ (columns 3 and 4), and $10 \rm G_0$ (columns 5 and 6). The left-hand panels of each set are scattered plots of the final mass and semi-major axis of all bodies (planets and embryos) remaining at the end of the simulation with increasing shielding time $t_{\textrm{sh}}$ from top to bottom. The right-hand panels show the number density heatmap of all bodies included in the corresponding left-hand panels, with darker shade indicating more bodies populated in that region of final mass and semi-major axis. 

\FloatBarrier

\bsp	
\label{lastpage}
\end{document}